\documentclass[12pt,a4paper,english,amssymb,nofootinbib,superscriptaddress]{revtex4}
\usepackage[caption=false,position=top]{subfig}
\usepackage{amsmath}
\usepackage{mathtools}
\usepackage{cancel}
\usepackage{enumitem}
\usepackage{amssymb}
\usepackage{graphicx}
\usepackage{tikz}
\usepackage{tkz-tab}
\usepackage{xpatch}
\usepackage{color}
\usepackage{verbatim}
\usepackage{amsmath}
\usepackage{amssymb}
\usepackage{esint}
\usepackage{float}
\usepackage[unicode=true,pdfusetitle,bookmarks=true,bookmarksnumbered=false,bookmarksopen=false,breaklinks=false,pdfborder={0 0 1},backref=false,colorlinks=true]{hyperref}
\hypersetup{citecolor=blue,urlcolor=blue,linkcolor=blue}

\makeatletter

\pdfpageheight\paperheight
\pdfpagewidth\paperwidth

\@ifundefined{textcolor}{}
{%
 \definecolor{BLACK}{gray}{0}
 \definecolor{WHITE}{gray}{1}
 \definecolor{RED}{rgb}{1,0,0}
 \definecolor{GREEN}{rgb}{0,1,0}
 \definecolor{BLUE}{rgb}{0,0,1}
 \definecolor{CYAN}{cmyk}{1,0,0,0}
 \definecolor{MAGENTA}{cmyk}{0,1,0,0}
 \definecolor{YELLOW}{cmyk}{0,0,1,0}
}

\@ifundefined{definecolor}{color}{}
\usepackage{babel}
\pdfpageheight\paperheight
\pdfpagewidth\paperwidth

\@ifundefined{textcolor}{}{%
 \definecolor{BLACK}{gray}{0}
 \definecolor{WHITE}{gray}{1}
 \definecolor{RED}{rgb}{1,0,0}
 \definecolor{GREEN}{rgb}{0,1,0}
 \definecolor{BLUE}{rgb}{0,0,1}
 \definecolor{CYAN}{cmyk}{1,0,0,0}
 \definecolor{MAGENTA}{cmyk}{0,1,0,0}
 \definecolor{YELLOW}{cmyk}{0,0,1,0}
 }

\@ifundefined{definecolor}{color}{}
\usepackage{latexsym}\usepackage{bm}

\usepackage{amssymb}

\makeatother

\begin{document}

\title{Orbits around a Kerr black hole and its shadow}

\author{Onur U\c{c}anok}

\email{ucanokonur@gmail.com}

\selectlanguage{english}%

\affiliation{Department of Physics,\\
 Middle East Technical University, 06800 Ankara, Turkey}

\begin{abstract}
Since the full \textit{General Theory of Relativity} has been unveiled to the scientific community in 1915, many solutions to the vacuum Einstein field equations have been found and studied \cite{stephani_2003}. This paper aims at documenting exhaustively the derivation of the shape of the patch of the sky that is left completely black by a spinning black hole described by the \textit{Kerr} solution in \textit{Boyer-Lindquist} coordinates. This dark zone in the observer's sky is called the \textit{black hole shadow}. Conserved quantities that allow for the analysis of particle orbits are first introduced, with the help of which the trajectories of photons are uniquely described by two impact parameters (Specific angular momentum in the azimuthal direction -$\mathcal L$- and dimensionless Carter's constant -$\mathcal Q^\circ$-). We then derive the conditions on those parameters required for a photon to be captured by the black hole. These conditions are then translated into an equation for the black hole shadow. We conclude the paper by drawing out the black hole shadows for an equatorial observer for two separate cases.

\tableofcontents{}
\[
\]

\end{abstract}

\maketitle

\section{Introduction}
The Kerr solution to the Einstein Vacuum Field Equations was first introduced by Roy Patrick Kerr in 1963 \cite{kerr_1963,teukolsky_2015}. This solution came after many failed attempts from notable scientists (such as Lewis and Papapetrou) to crack the equations into an exact and asymptotically flat description of spacetime outside a rotating object \cite{dautcourt_2008}. It was originally discovered in \textit{Eddington-Finkelstein} coordinates -named after the Eddington-Finkelstein coordinates of the Schwarzschild metric it reduced to for the non-rotating limit-. Later simplified into the \textit{Boyer-Lindquist} coordinates in 1967, this new coordinate gave a better insight into the important surfaces of the metric, along with an easy "conversion" of the metric to its charged counterpart : the \textit{Kerr-Newman} metric. In these coordinates, the line element of the metric reads
\begin{equation}
\begin{split}
ds^2_{\text{Kerr}} & = \overbrace{- (1-\frac{2Mr}{\Sigma})}^{g_{tt}} dt^2 - \overbrace{\frac{4Mra \sin^2 \theta}{\Sigma}}^{2g_{t \phi}} dt d \phi +\overbrace{\frac{\Sigma}{\Delta}}^{g_{rr}} dr^2 \\ 
& \qquad \qquad \qquad + \overbrace{\Sigma}^{g_{\theta \theta}} d\theta ^2 + \overbrace{(r^2 + a^2 + \frac{2Ma^2 r\sin^2\theta}{\Sigma})\sin^2\theta}^{g_{\phi \phi}} d\phi^2.
\end{split}
\end{equation}
With $a \equiv \frac{J}{M}$, $\Sigma \equiv r^2 + a^2 \cos^2 \theta$ and $\Delta \equiv r^2 - 2Mr + a^2$. \newline
Or equivalently, since $ds^2 = g_{\mu \nu} dx^\mu dx^\nu$, expressing $g_{\mu \nu}$ and its inverse $g^{\mu \nu}$ in matrix form
\begin{equation}
g_{\mu \nu} = \begin{bmatrix}
    - (1-\frac{2Mr}{\Sigma}) & 0 & 0 & - \frac{2Mra \sin^2 \theta}{\Sigma} \\
    0 & \frac{\Sigma}{\Delta} & 0 & 0 \\
     0 & 0 & \Sigma & 0 \\
    - \frac{2Mra \sin^2 \theta}{\Sigma} & 0 & 0 & (r^2 + a^2 + \frac{2Mra^2 \sin^2\theta}{\Sigma})\sin^2\theta
\end{bmatrix}.
\end{equation}
\begin{equation}
g^{\mu \nu} = \begin{bmatrix}
    - \frac{1}{\Delta}(r^2+a^2+\frac{2Mra^2\sin^2\theta}{\Sigma}) & 0 & 0 & -\frac{2Mra}{\Delta \Sigma} \\
    0 & \frac{\Delta}{\Sigma} & 0 & 0 \\
     0 & 0 & \frac{1}{\Sigma} & 0 \\
    -\frac{2Mra}{\Delta \Sigma} & 0 & 0 & \frac{\Delta- a^2 \sin^2\theta}{\Delta \Sigma \sin^2\theta}
\end{bmatrix}.
\end{equation}
It can be seen to be azimuthally symmetric - metric independent on the azimuthal angle $\phi$- and static - metric independent on the coordinate time $t$ -. This solution to the vacuum field equation is associated with the spacetime around -not inside- a rotating body of mass $M$ and angular momentum $J$. \cite{frolov_novikov_1998} \newline
In the following few sections of this paper, we will study the infinite redshift surfaces, followed by the event horizon associated with this metric. \newline
We follow up by introducing the reader to conserved quantities for a free particle inside this spacetime, which are used in the expression of the equations governing the trajectories of these particles. These equations will give us some foothold onto which we shall build the understanding that photon orbits are uniquely defined by the conserved charges defined previously, most notably upon the azimuthal angular momentum $\mathcal L$ and the Carter's constant $\mathcal Q^\circ$. \newline
We notice that the parameters required by a photon to either escape or get captured by the black hole are separated by some critical impact parameters $\mathcal L_c$ and $\mathcal Q_c^\circ$, represented by a curve in the parameter space, which we draw for visualization. \newline
We then initiate the reader with \textit{Celestial Coordinates}: coordinates defined in the sky dome of the observer. For an observer sufficiently far from the black hole, we convert the critical parameters to the sky of the observer, such that we get an equation of the outlines of the shadow of this black hole. \newline
We finally finish things off by considering an equatorial observer looking at the black hole, and draw the outlines of the dark region of its sky for both slowly rotating black holes ($\alpha \ll 1$) and extremal black holes. ($\alpha \approx 1$)\newline
The main objective of this paper has been to create an understandable and comprehensive derivation of the process of finding a black hole shadow associated with a rotating black hole.

\section{Infinite redshift surface}
Let us investigate the redshift of a photon propagating radially out of this black hole. 
\begin{figure}[H]
\centering
\begin{tikzpicture}
\filldraw[color=black!5, fill=black!40, very thick](-5,0) circle (1.5);
\filldraw [thick, gray] (0,0) circle (2pt) node[anchor=south] {$p^\mu = (\omega,\vec p,0,0)$};
\filldraw [thick, gray] (5,0) circle (2pt) node[anchor=south] {$p'^\mu = (\omega',\vec p,0,0)$};
\draw[thick, ->] (0,0)--(1,0);
\draw[thick, ->] (5,0)--(6,0);
\draw[<->] (-5,-1)--(0,-1) node[anchor=south] {$r$} ;
\draw[<->] (-5,-2)--(5,-2)  node[anchor=south] {$r'$} ;
\end{tikzpicture}
\caption{Free photon propagating radially outwards. The photon at $r$ is propagated to $r'$. Since no external forces are present, $\vec{p}$ is untouched.}
\end{figure}
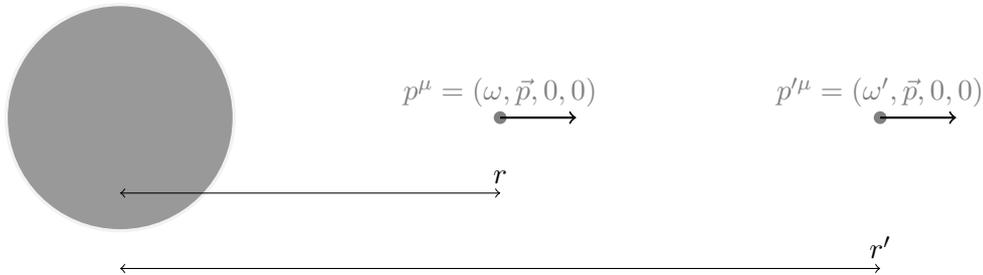
Knowing that for an affinely parametrized photon momentum, we have $p^\mu p_\mu = 0$ \cite{carroll_2014}.
\begin{equation}
\begin{split}
p^\mu p_\mu & = 0, \\
-(1-\frac{2Mr}{\Sigma}) \omega^2 + \frac{\Sigma}{\Delta} \vec p^{\ 2} & = 0. \\
p'^\mu p'_\mu & = 0, \\
-(1-\frac{2Mr'}{\Sigma'}) \omega'^2 + \frac{\Sigma'}{\Delta'} \vec p^{\ 2} & = 0. \\
\end{split}
\end{equation}
For a free particle, the spatial component of the momentum $\vec{p}$ does not change under translation. Using this, we equate the $\vec{p}^{\ 2}$ terms, getting a relation between the frequencies at these different points
\begin{equation}
\frac{\Delta}{\Sigma}(1-\frac{2Mr}{\Sigma}) \omega^2 = \frac{\Delta'}{\Sigma'}(1-\frac{2Mr'}{\Sigma'}) \omega'^2.
\end{equation}
In other words, for a photon released on the surface $I$ on which $g_{tt}|_I = 0$, the redshift further away becomes so large that the signal is infinitely redshifted ($\omega' = 0$). This surface is called the \textit{infinite redshift surface}, the equation of which can be found;
\begin{equation}
g_{tt}|_I = - (1-\frac{2Mr}{\Sigma}) = \frac{- r^2 - a^2 \cos ^2 \theta + 2Mr}{\Sigma} = 0.
\end{equation}
The denominator is never singular, so this equation for the surface is well defined everywhere. Solving the second degree polynomial in the numerator, we get the inner and outer infinite redshift surfaces $r_{\text{rs}}$ for the black hole
\begin{equation}
\label{rirpm}
r_{\text{rs}} = M \pm \sqrt{M^2 - a^2 \cos ^2 \theta}.
\end{equation}

\begin{figure}[H]
\centering
\subfloat[][$\alpha = 0$]{
  \includegraphics[width=0.3\columnwidth]{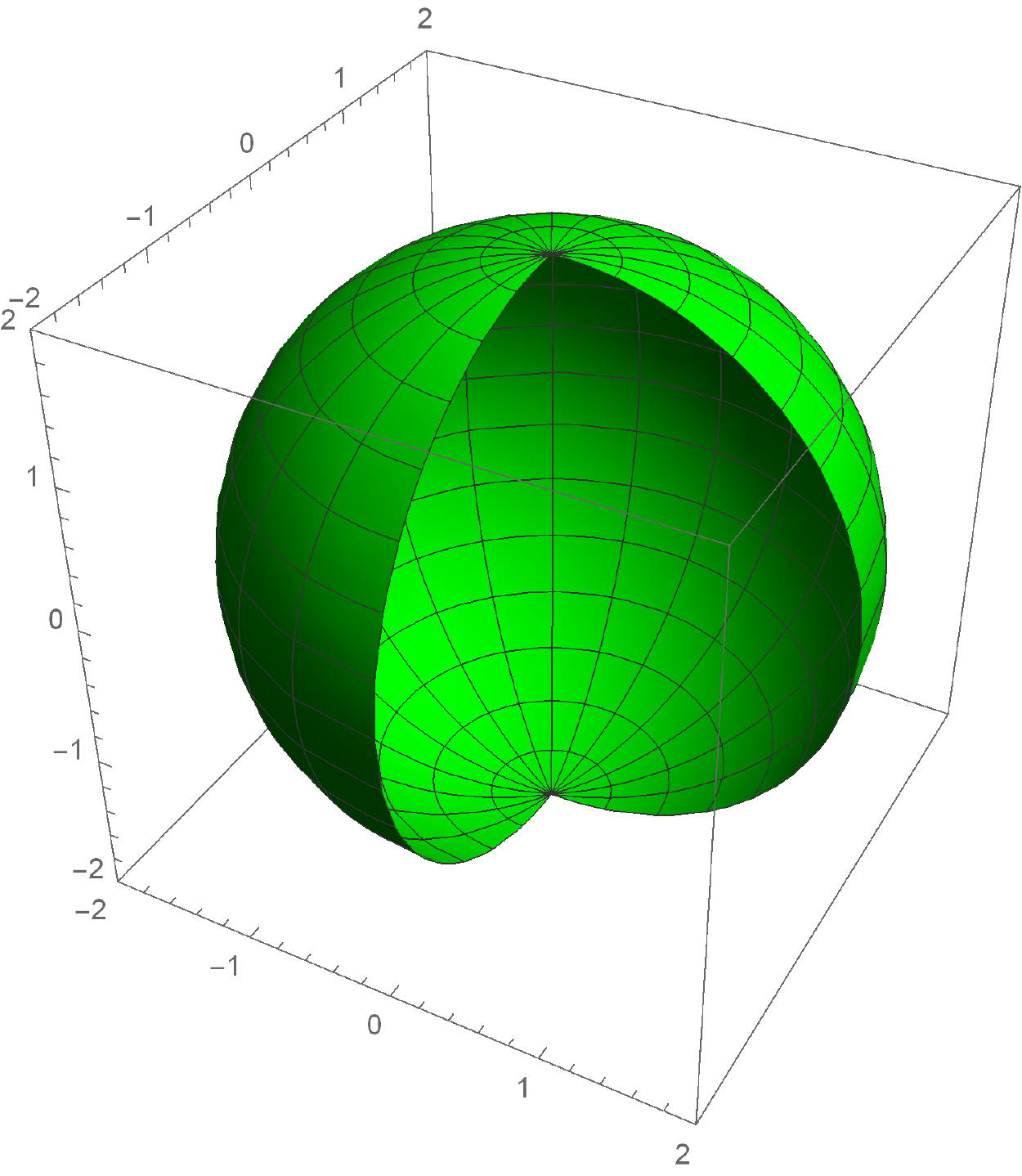}
}\hspace{3cm}
\subfloat[][$\alpha = 0.33$]{
  \includegraphics[width=0.3\columnwidth]{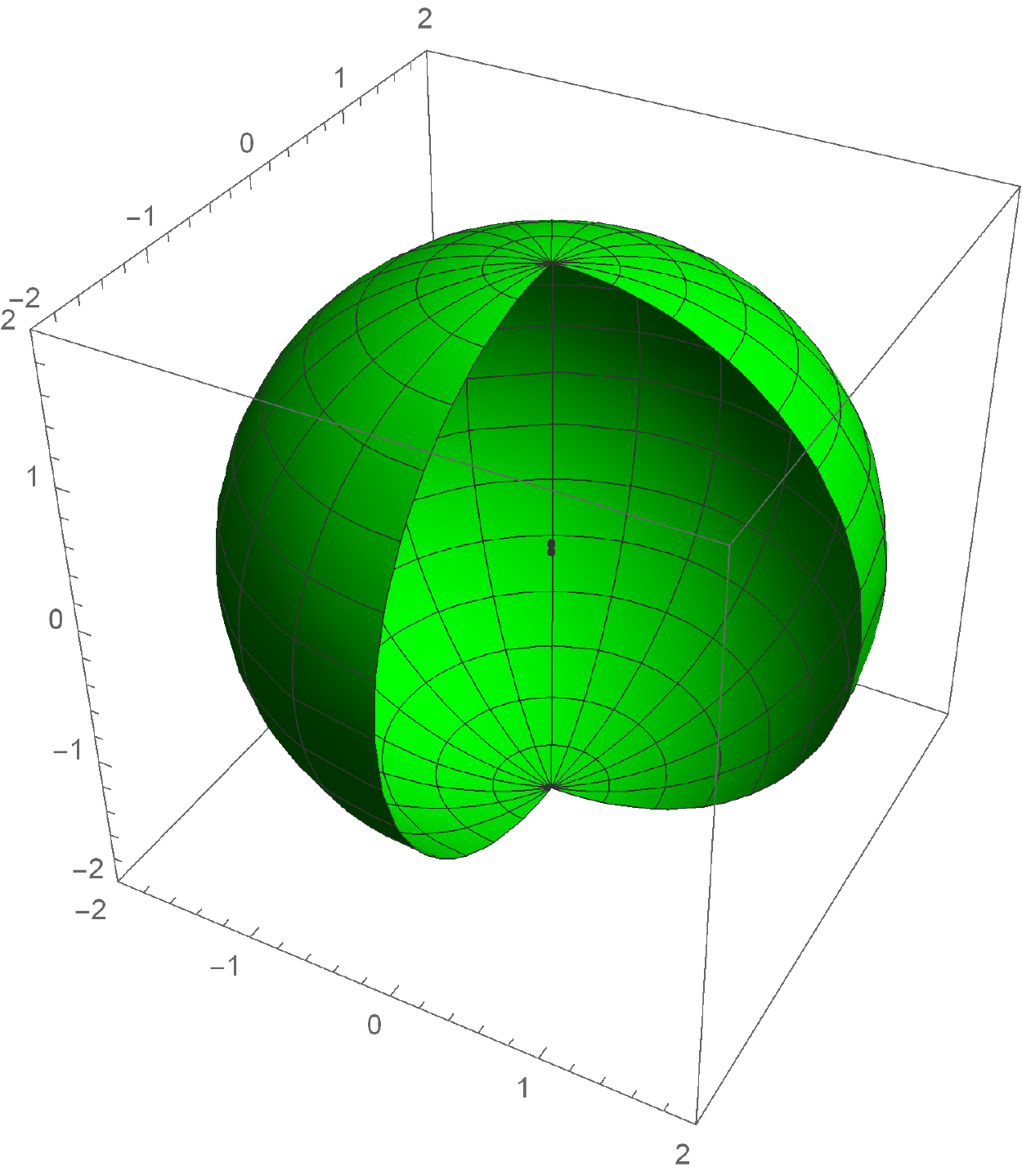}
}
\end{figure}
\begin{figure}[H]
\centering
\subfloat[][$\alpha = 0.67$]{
  \includegraphics[width=0.3\columnwidth]{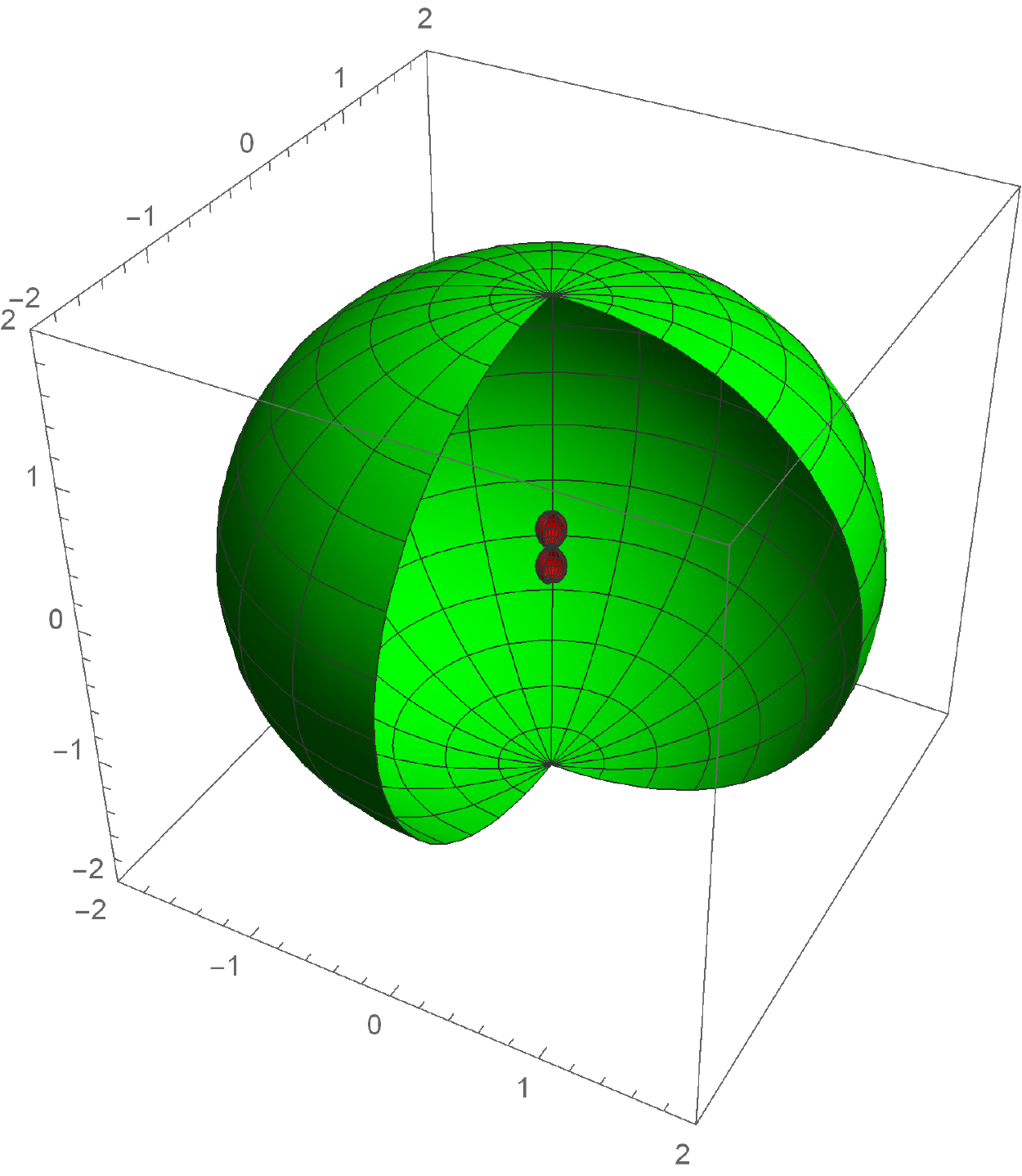}
}\hspace{3cm}
\subfloat[][$\alpha = 1$]{
  \includegraphics[width=0.3\columnwidth]{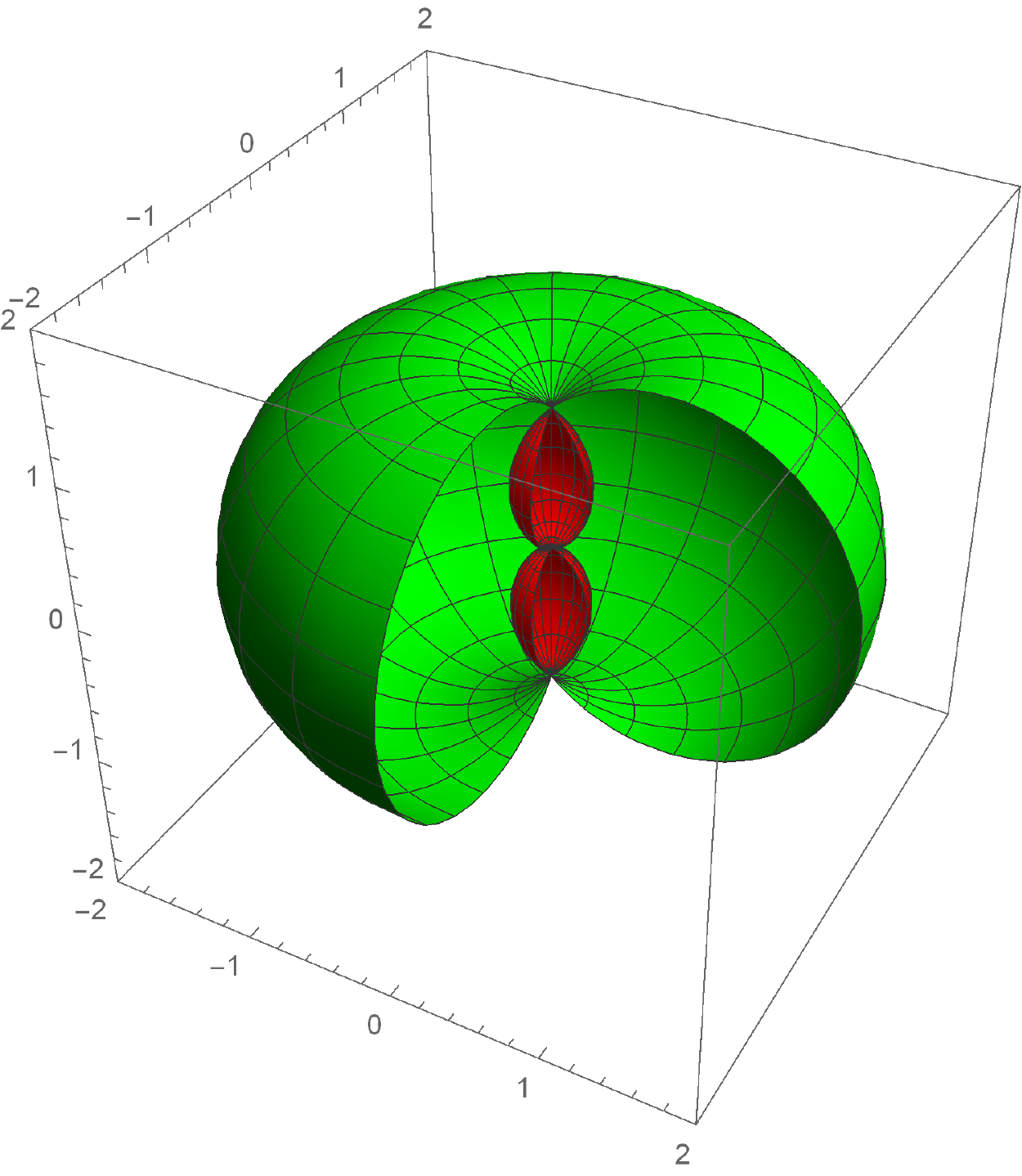}
}

\caption{Inner (red) and outer (green) infinite redshift surfaces drawn for various spin parameter $\alpha = \frac{a}{M}$ of the black hole}
\end{figure}

It is tempting to surmise that this surface is the event horizon of the black hole -the point after which nothing can return-. After all, light gets infinitely redshifted, so does that not mean that it cannot escape the black hole after that point ? To get further insight into this surface's properties, let us denote it's normal $n_\mu = \partial_\mu (r-r_\text{ir})$. Writing explicitly all of its components to find its norm :
\begin{equation}
\begin{split}
n_\mu & = (0,1,\pm \frac{a^2 \sin \theta \cos \theta}{\sqrt{M^2-a^2 \cos^2 \theta}},0). \\
n_\mu n^\mu |_{I}& = g^{rr}+ g^{\theta \theta} \frac{a^4 \sin^2 \theta \cos^2 \theta}{M^2-a^2 \cos^2 \theta} = \frac{\Delta}{\Sigma} +\frac{a^4 \sin^2 \theta \cos^2 \theta}{\Sigma (M^2-a^2 \cos^2 \theta)} \\
& = \frac{(M^2-a^2 \cos^2 \theta)\overbrace{(r^2-2Mr+a^2)}^{= a^2 \sin^2 \theta} + a^4 \sin ^2 \theta \cos^2 \theta}{\Sigma (M^2-a^2 \cos^2 \theta)} \\
& = \frac{M^2 a^2 \sin^2 \theta}{\Sigma (M^2-a^2 \cos^2 \theta)}.
\end{split}
\end{equation}
So for $a<M$, it is always true that both of these surfaces have spacelike normal vectors. This implies the surfaces themselves are merely timelike -very much roamable by causal creatures such as us-.

\section{The event horizon}
The critical surface $\Delta = 0$ seems like a good second candidate for an event horizon. The surface's equation in terms of the coordinates can be found as follows
\begin{equation}
\begin{split}
\Delta & = r^2 - 2Mr + a^2= 0. \\
r_\pm & = M \pm \sqrt{M^2 - a^2}.
\end{split}
\end{equation}
With a normal vector $n_\mu = \partial_\mu (r - r_\pm)$, when expressed term by term, we can find its norm.
\begin{equation}
\begin{split}
n_\mu & = (0,1,0,0). \\
n^\mu n_\mu & = g^{rr}|_{\Delta = 0} = \frac{\Delta}{\Sigma}|_{\Delta = 0}= 0.\\
\end{split}
\end{equation}
This surface is indeed null, as we expect an event horizon to be.

\begin{figure}[H]
\centering
\subfloat[][$\alpha = 0$]{
  \includegraphics[width=0.3\columnwidth]{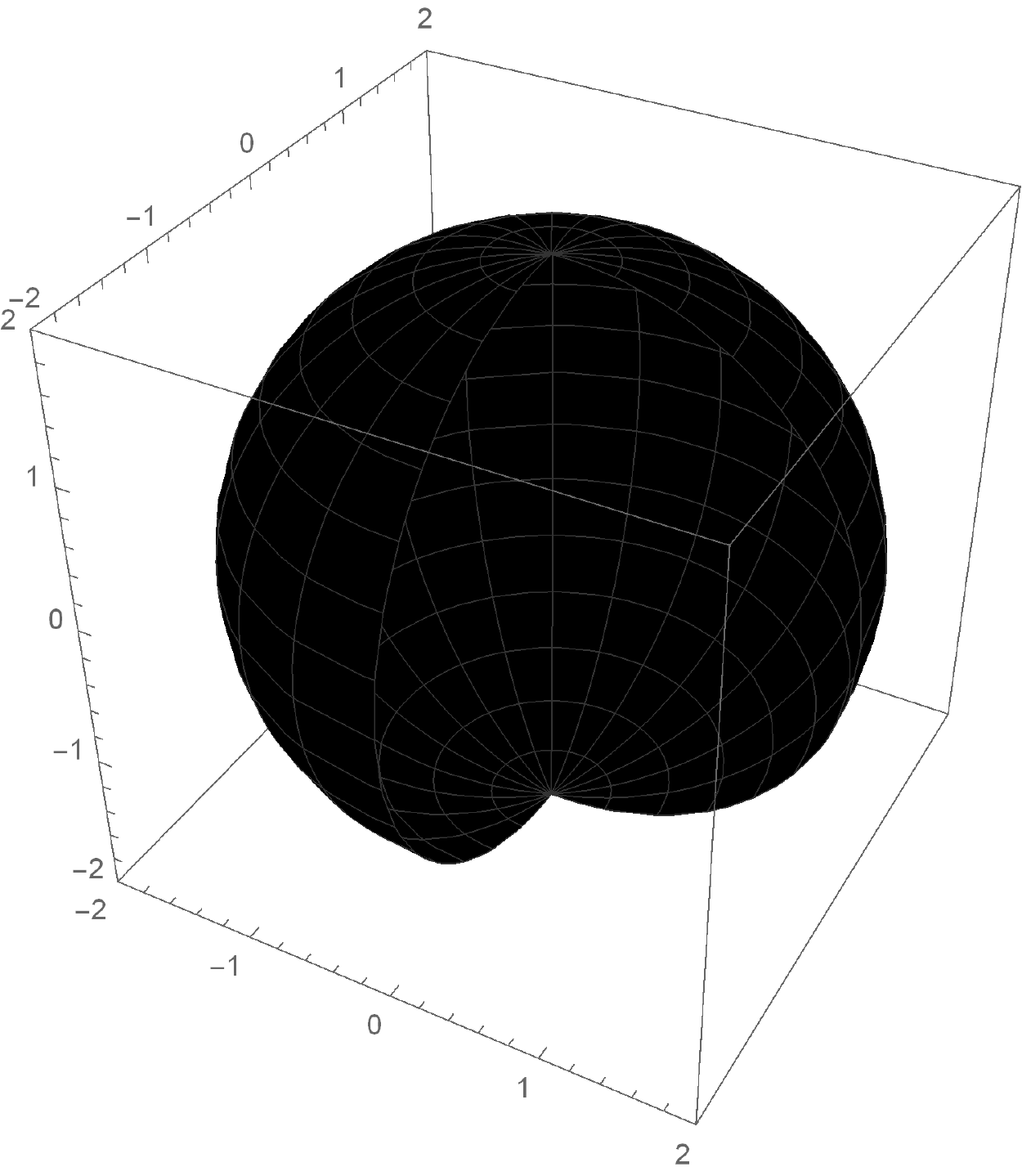}
}\hspace{3cm}
\subfloat[][$\alpha = 0.33$]{
  \includegraphics[width=0.3\columnwidth]{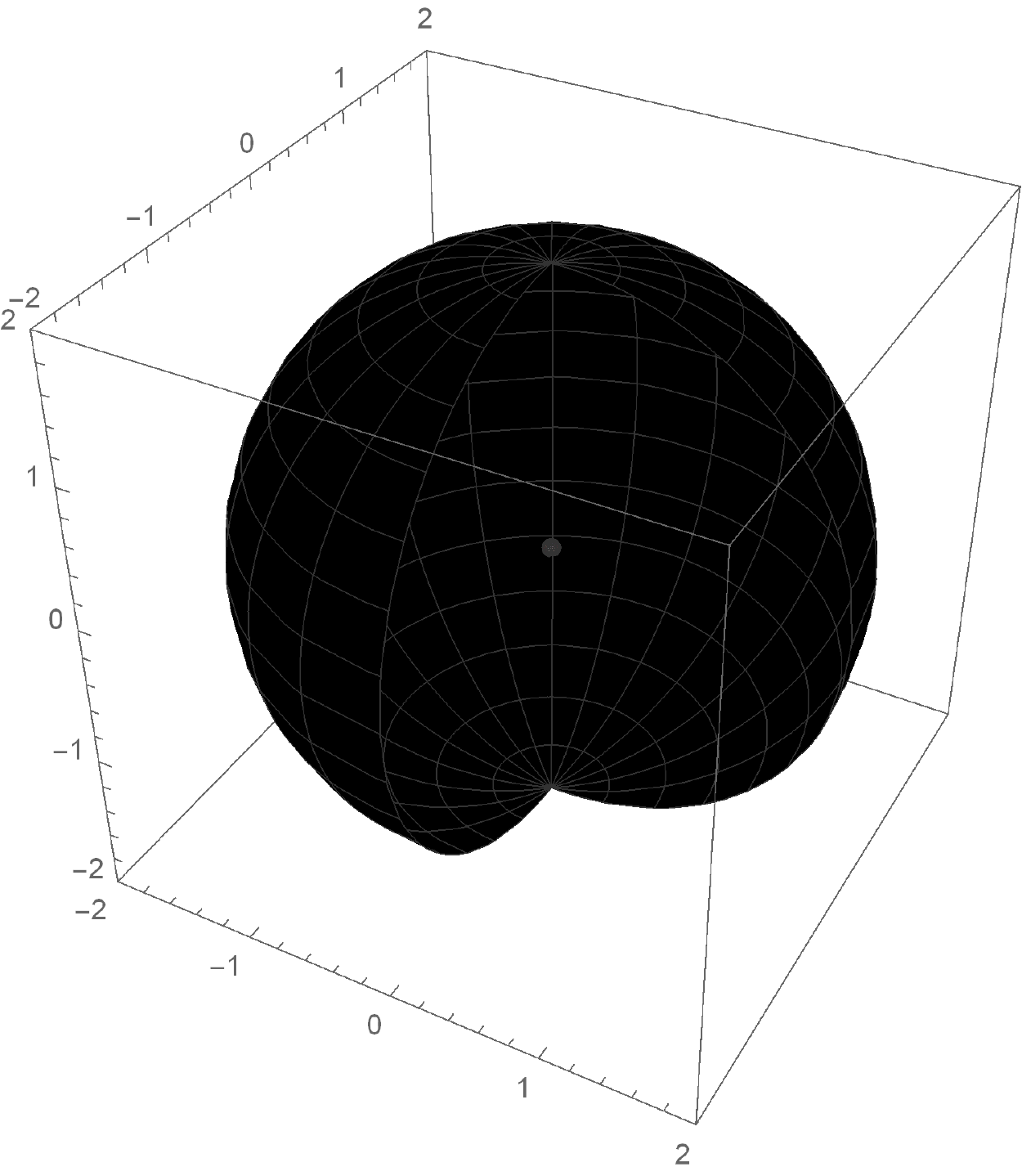}
}
\end{figure}
\begin{figure}[H]
\centering
\subfloat[][$\alpha = 0.67$]{
  \includegraphics[width=0.3\columnwidth]{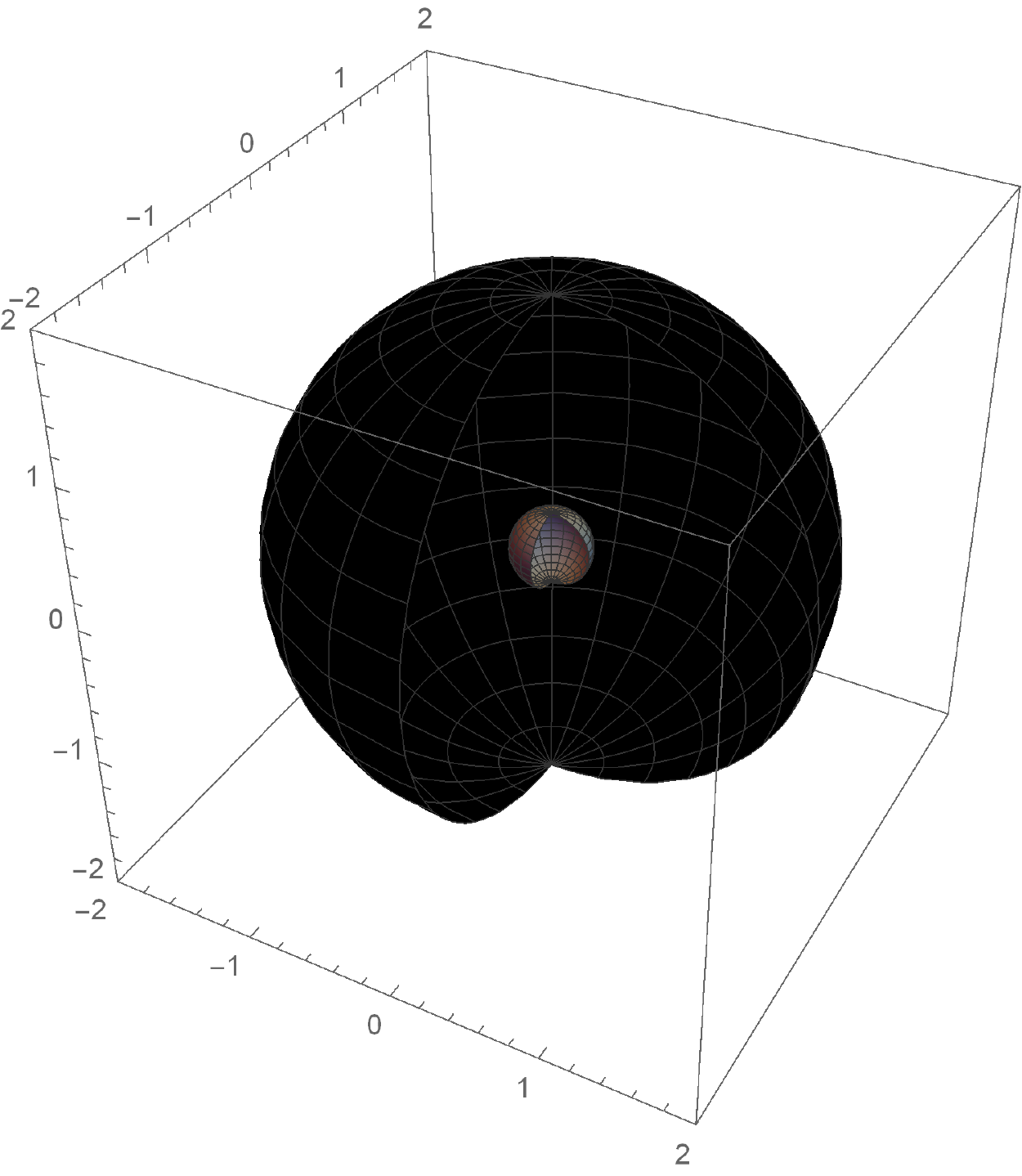}
}\hspace{3cm}
\subfloat[][$\alpha = 1$]{
  \includegraphics[width=0.3\columnwidth]{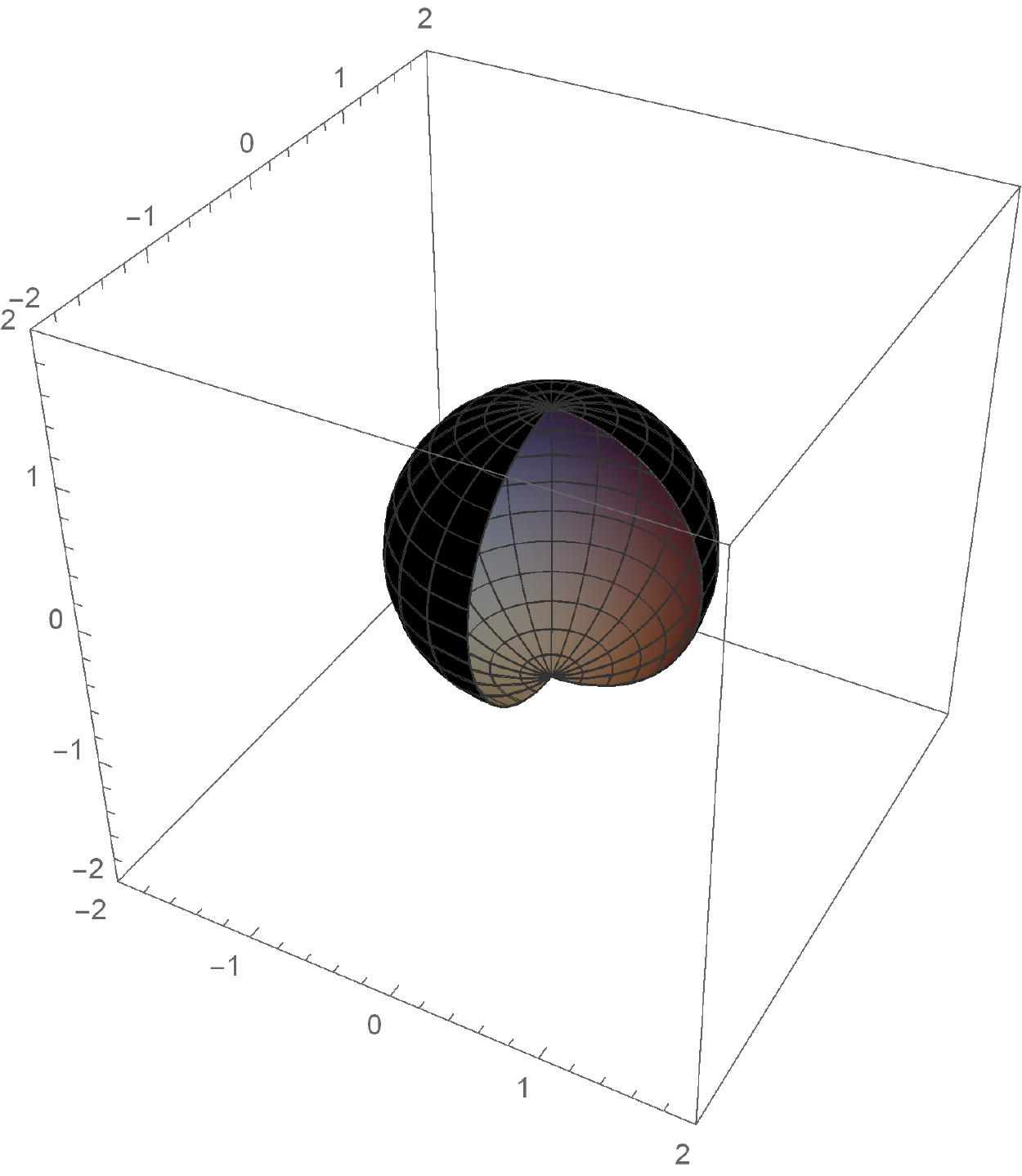}
}

\caption{Inner (gray) and outer (black) event horizon surfaces drawn for various spin parameter $\alpha$ of the black hole}

\end{figure}

Assume an observer in the equatorial plane ($\theta = \frac{\pi}{2}$ and $\dot \theta = 0$) doing circular motion ($r = r_c$ and $\dot r = 0$). Its four velocity can be taken as
\begin{equation}
u^\mu = \frac{\eta^\mu}{(\eta^2)^{\frac{1}{2}}} \qquad \qquad \text{Where } \eta^\mu = (1,0,0,\omega).
\end{equation}
This vector has different norms for different $\omega$ values :
\begin{equation}
\eta^\mu \eta_\mu =  g_{tt} + 2 g_{t \phi} \omega + g_{\phi \phi} \omega^2.
\end{equation}
The roots of the above second degree polynomical are the points when the sign of the norm of $u^\mu$ changes.
\begin{equation}
\omega_\pm = \frac{- g_{t \phi} \pm \sqrt{g_{t \phi}^2 - g_{tt} g_{\phi \phi}}}{g_{\phi \phi}}.
\end{equation}
We should expand on the term in the square root, since that is going to determine the nature of the roots.
\begin{equation}
\begin{split}
g_{t \phi}^2 - g_{tt} g_{\phi \phi} & = \frac{4 M^2 r^2 a^2 \sin ^4 \theta}{\Sigma^2} + (1-\frac{2Mr}{\Sigma})(r^2 + a^2 + \frac{2Mr a^2 \sin^2 \theta}{\Sigma}) \sin^2 \theta \\
& = \big( \cancelto{i}{\frac{4 M^2 r^2 a^2 \sin^2 \theta}{\Sigma^2}} + r^2 + a^2 + \frac{2Mr a^2 \sin^2 \theta}{\Sigma} \\
& \qquad \qquad \qquad \qquad - \frac{2Mr^3}{\Sigma} -\frac{2Mra^2}{\Sigma} \cancelto{i}{- \frac{4M^2r^2 a^2 \sin^2 \theta}{\Sigma}} \big) \sin^2 \theta \\
& = \big( r^2 + a^2 - 2Mr\frac{r^2 + a^2 - a^2 \sin ^2 \theta}{\Sigma} \big) \sin^2 \theta \\
& = \big( \underbrace{r^2 + a^2 - 2Mr\frac{\overbrace{r^2 + a^2 \cos ^2 \theta}^{= \Sigma}}{\Sigma} }_{= \Delta}\big) \sin^2 \theta \\
& = \Delta \sin^2\theta.
\end{split}
\end{equation}
Since $ 0 \leq \sin^2 \theta \leq 1 $ for real $\theta$, we can classify the possible sign of the norm of $u^\mu$ as follows:
\begin{enumerate}
\item When $\Delta > 0$ (outside the event horizon), the roots are distinct
\begin{center}
\begin{tikzpicture}
\tkzTabInit[lgt=2,espcl=2,deltacl=0]
  {$\omega$ /.8, $u^\mu u_\mu$ /.8}
  {,$\omega_-$,$\omega_+$,}
\tkzTabLine {,+,z,-,z,+,}
\end{tikzpicture}
\end{center}
There are values of $\omega$ for which the motion is timelike. We conclude that particles can be made to enter circular motion outside the event horizon. One thing to note is that inside the infinite redshift surface, we have $g_{tt} > 0$ which implies that $\omega_-$ is positive valued ($\omega_+$ is positive, too). This, in turn, implies that particles inside the \textit{ergosphere} (The region of the manifold between the event horizon and the infinite redshift surface) can only be corotating with the black hole.
\item When $\Delta	= 0$, the roots are equal
\begin{center}
\begin{tikzpicture}
\tkzTabInit[lgt=2,espcl=2,deltacl=0]
  {$\omega$ /.8, $u^\mu u_\mu$ /.8}
  {,$\omega_\pm$,}
\tkzTabLine {,+,z,+,}
\end{tikzpicture}
\end{center}
The motion can never be timelike for any $\omega_\pm$. So the only equatorially circular motion on the event horizon can be achieved by massless particle. Please note that this analysis doesn't assume particles on geodesics. In this case, a free massless particle cannot do a circular motion on top of the event horizon. Such a motion is only possible in the presence of some external force.
\item When $\Delta < 0$, the roots are imaginary
\begin{center}
\begin{tikzpicture}
\tkzTabInit[lgt=2,espcl=2,deltacl=0]
  {$\omega$ /.8, $u^\mu u_\mu$ /.8}
  {,}
\tkzTabLine {,+,}
\end{tikzpicture}
\end{center}
The motion is spacelike for all values of $\omega$. Even with external force, there is no way for a particle to enter circular motion inside the event horizon.
\end{enumerate}

\section{Equations of motion for an arbitrary particle}
One might be tempted to use the geodesic equations to find the equations of motion generally for an arbitrary particle. Although this seems feasible at first, writing down these equations gives us a multitude of coupled second degree non linear differential equations, from which it is hard to derive the real physics at play \cite{vazquez_esteban_2018}. To remedy this, we need a couple of specifications about the orbiting particle; conserved charges, along with Hamilton-Jacobi Theory. \newline 
In the coordinates  $(t,r,\theta,\phi)$ given above, the Killing vector field corresponding to the field being static is $K_{t}^\mu = (1, 0, 0, 0)$ while the Killing vector field corresponding to the azimuthal symmetry is $K_{\phi}^\mu = (0,0,0,1)$. Conserved charges associated with these are (See Appendix C \& D)
\begin{equation}
\begin{split}
E & = - p_\mu K_t^\mu = - g_{\mu \nu} p^\mu K_t^\nu = - g_{tt} \ \dot{t} - g_{t \phi} \ \dot{\phi} \\
& = (1-\frac{2Mr}{\Sigma}) \dot{t} + \frac{2Mra \sin^2 \theta}{\Sigma} \dot{\phi} \ . \\
L & = p_\mu K_\phi^\mu =  g_{\mu \nu} p^\mu K_\phi^\nu = g_{t\phi} \dot{t} + g_{\phi \phi} \dot{\phi} \\
& = - \frac{2Mra \sin^2 \theta}{\Sigma} \ \dot{t} + \big( r^2 + a^2 + \frac{2Mr a^2 \sin^2 \theta}{\Sigma} \big) \sin^2 \theta \ \dot{\phi} \ .
\end{split}
\end{equation}
Where the dot denotes derivative with respect to the proper time $\tau$ chosen for our desired particle's study. ($ \frac{d}{d \tau} = \dot{ \  }$). Another thing to note here is that $L$ is strictly the \textit{azimuthal} component of the angular momentum of the particle. \newline
Consider the lagrangian for this particle :
\begin{equation}
\mathcal{L}(x^\mu, \dot{x}^\mu) = \frac{m}{2} g_{\mu \nu} \dot{x}^\mu \dot{x}^\nu.
\end{equation}
The conjugate canonical four-momenta can be defined as
\begin{equation}
p_\mu = \frac{\partial \mathcal{L}}{\partial \dot{x^\mu}} = m g_{\mu \nu} \dot{x^\nu} \rightarrow \dot{x}^\mu(p_\alpha) = \frac{1}{m} g^{\mu \nu} p_\nu .
\end{equation}
And the hamiltonian is defined as
\begin{equation}
\begin{split}
\mathcal{H}(x^\alpha, p_\alpha) & = p_\nu \dot{x}^\nu (p_\alpha)- \mathcal{L}(x^\mu,\dot{x}^\mu(p_\alpha)) = \frac{1}{m} g^{\mu \nu} p_\nu p_\mu - \frac{m}{2} g_{\mu \nu} \dot{x}^\mu(p_\alpha) \dot{x}^\nu(p_\alpha) \\
& = p^\mu p_\mu - \frac{1}{2m} \overbrace{g_{\mu \nu} g^{\mu \alpha}}^{\delta_\nu^{\ \alpha}} g^{\nu \beta} p_\alpha p_\beta = \frac{1}{2m} p^\mu p_\mu.
\end{split}
\end{equation}
Consider now a transformation which preserves the first degree derivative nature of the hamiltonian equations $\frac{\partial K}{\partial x^\alpha}=- \dot{p}^\alpha$ and $\frac{\partial K}{\partial p^\alpha}= \dot{x}^\alpha$ (Where $K$ represent the new hamiltonian in this new coordinate). These are called canonical transformation \cite{goldstein_poole_safko_2014}. The generating function $U = U(x^\mu, \tau)$ defined as the function which generates the new hamiltonian $K$ from the old $H$ with the following Hamilton-Jacobi equation $K = H(x^\mu, \frac{\partial U}{\partial x^\mu}) + \frac{\partial U}{\partial \tau} = 0$. This generating function $U$ satisfies $\frac{\partial U}{\partial x^\mu} = p_\mu$ by construction. \newline
The conserved charges we've found above, coupled with our $\frac{\partial U}{\partial x^\mu} = p_\mu$ and $\mathcal{H}(x^\mu, \frac{\partial U}{\partial x^\mu}) + \frac{\partial U}{\partial \tau} = 0$ conditions gives us the first form for our generating function $U(x^\mu, \tau)$ (See Appendix A)
\begin{equation}
U(x^\mu, \tau) = \frac{m}{2}\tau - Et + L\phi + U_{r \theta}(r,\theta).
\end{equation}
Where $U_{t \theta}(r,\theta)$ is some $r$ and $\theta$ dependent function which encodes the dynamics in those coordinates. We pick this function to be decomposable into $U_{r \theta}(r,\theta) = U_{r}(r) + U_{\theta}(\theta)$.
\begin{equation}
U(x^\mu, \tau) = \frac{m}{2}\tau - Et + L\phi + U_{r}(r) + U_{\theta}(\theta).
\end{equation}
Plugging the Hamiltonian equation $\mathcal{H}= \frac{1}{2m} p^\mu p_\mu$ into the Hamilton-Jacobi equation with $p^\mu p_\mu = - m^2$
\begin{equation}
\begin{split}
H(x^\mu, \frac{\partial U}{\partial x^\mu}) + \frac{\partial U}{\partial \tau} & = 0 \\
\frac{1}{2m} g^{\mu \nu} p_\nu p_\mu + \frac{m}{2} & = 0 \\
g^{t t} p_t p_t + 2 g^{t \phi} p_t p_\phi + g^{r r} p_r p_r + g^{\theta \theta} p_\theta p_\theta + g^{\phi \phi} p_\phi p_\phi + m^2 & = 0.
\end{split}
\end{equation}
For the momenta, we merely replace $p_t = -E$ and $p_\phi = L$ for the cyclic coordinates, while we write out the non-cyclic ones in terms of derivatives of $U_{r/\theta}$. Further simplifications then leads to
\begin{equation}
\begin{split}
g^{t t} E^2 - 2 g^{t \phi} EL + g^{r r} \left(\frac{\partial U_{r}(r)}{\partial r} \right)^2 + g^{\theta \theta} \left(\frac{\partial U_{\theta}(\theta)}{\partial \theta}\right)^2 + g^{\phi \phi} L^2 + m^2 & = 0 \\
- \frac{1}{\Delta}(r^2+a^2+\frac{2Mra^2\sin^2\theta}{\Sigma}) E^2 + \frac{4Mra}{\Delta \Sigma} EL + \frac{\Delta}{\Sigma} \left(\frac{\partial U_{r}(r)}{\partial r} \right)^2 + & \\
\qquad \qquad \qquad  \frac{1}{\Sigma} \left(\frac{\partial U_{\theta}(\theta)}{\partial \theta}\right)^2 + \frac{\Delta- a^2 \sin^2\theta}{\Delta \Sigma \sin^2\theta} L^2 + m^2 & = 0.
\end{split}
\end{equation}
Simplifying the $E^2$'s coefficient term
\begin{equation}
\begin{split}
r^2+a^2+\frac{2Mra^2\sin^2\theta}{\Sigma} & = \frac{(r^2+a^2)(r^2 + a^2 \cos^2 \theta) + 2Mra^2\sin^2\theta}{\Sigma} \\
& = \frac{(r^2+a^2)^2 - (r^2 + a^2) a^2 \sin^2\theta + 2Mra^2\sin^2\theta}{\Sigma} \\
& = \frac{(r^2+a^2)^2 -  a^2 \sin^2\theta \overbrace{(r^2 + a^2 - 2Mr)}^{\Delta}}{\Sigma} \\
& = \frac{(r^2+a^2)^2 -  a^2 \sin^2\theta \Delta}{\Sigma}.
\end{split}
\end{equation}
Using the fact that $\Sigma \neq \infty$, we get rid of the denominator. The Hamilton-Jacobi equation becomes
\begin{equation}
\begin{split}
(-\frac{(r^2+a^2)^2}{\Delta} + a^2 \sin^2\theta) E^2 + \frac{4Mra}{\Delta} EL + \Delta \left(\frac{\partial U_{r}(r)}{\partial r} \right)^2 + \left(\frac{\partial U_{\theta}(\theta)}{\partial \theta}\right)^2 & \\
+ (\frac{1}{\sin^2\theta} - \frac{a^2}{\Delta}) L^2 + (r^2 + a^2 \cos^2\theta) m^2 & = 0.
\end{split}
\end{equation}
This equation can be split into two separate equations with a separation constant $\mathcal{C}$ which will soon be related to another quantity $\mathcal Q$ called \textit{Carter's constant}.
\begin{equation}
\begin{split}
\left(\frac{\partial U_{\theta}(\theta)}{\partial \theta}\right)^2 & + a^2 \sin^2\theta E^2 + \frac{L^2}{\sin^2\theta} + m^2 a^2 \cos^2\theta = \\
& - \Delta \left(\frac{\partial U_{r}(r)}{\partial r} \right)^2 + \frac{(r^2+a^2)^2}{\Delta} E^2 - \frac{4Mra}{\Delta} EL + \frac{a^2}{\Delta} L^2 - m^2 r^2.
\end{split}
\end{equation}
Substracting a $2aEL$ term to both sides
\begin{equation}
\left(\frac{\partial U_{\theta}(\theta)}{\partial \theta}\right)^2 + \big( a\sin \theta E - \frac{L}{\sin \theta}\big)^2 + m^2 a^2 \cos^2\theta =  - \Delta \left(\frac{\partial U_{r}(r)}{\partial r} \right)^2 + \frac{\big( (r^2+ a^2)E - aL\big)^2}{\Delta} - m^2 r^2.
\end{equation}
Both sides of the equation are dependent on only one of two independent variables $r$ and $\theta$. They are thus equal to some constant $\mathcal{C}$
\begin{equation}
\begin{split}
\Delta \left(\frac{\partial U_{r}(r)}{\partial r} \right)^2 - \frac{\big( (r^2+ a^2)E - aL\big)^2}{\Delta} + m^2 r^2 & = - \mathcal{C}, \\
\left(\frac{\partial U_{\theta}(\theta)}{\partial \theta}\right)^2 + \big( a\sin \theta E - \frac{L}{\sin \theta}\big)^2 + m^2 a^2 \cos^2\theta & = \mathcal{C}.
\end{split}
\end{equation}
Which we can rewrite as
\begin{equation}
\begin{split}
\label{RTEOM}
\left(\frac{\partial U_{r}}{\partial r} \right)^2 & = \frac{\mathcal R(r)}{\Delta^2} \qquad \text{with } \mathcal R (r) \equiv \big( (r^2+ a^2)E - aL\big)^2 - \Delta \big(m^2 r^2 + \mathcal{C}\big) , \\
\left(\frac{\partial U_{\theta}}{\partial \theta}\right)^2 & = \Theta(\theta)  \qquad \text{with } \Theta(\theta) \equiv \mathcal C -  \big( a\sin \theta E - \frac{L}{\sin \theta}\big)^2 - m^2 a^2 \cos^2\theta.
\end{split}
\end{equation}
Or replacing $\frac{\partial U_r}{\partial r}$ and $\frac{\partial U_\theta}{\partial \theta}$ with $p_r$ and $p_\theta$ respectively
\begin{equation}
\begin{split}
\frac{\partial U_r}{\partial r} = p_r & = \frac{\partial \mathcal{L}}{\partial \dot{r}} = \frac{\partial\left(\frac{m}{2}g_{rr}\dot{r}\dot{r}\right)}{\partial \dot{r}} = \frac{\Sigma}{\Delta} m \dot{r}, \\
\frac{\partial U_\theta}{\partial \theta} = p_\theta & = \frac{\partial \mathcal{L}}{\partial \dot{\theta}} = \frac{\partial\left(\frac{m}{2}g_{\theta}\dot{\theta}\dot{\theta}\right)}{\partial \dot{\theta}} = \Sigma m \dot{\theta}.
\end{split}
\end{equation}
Plugging back into equation \eqref{RTEOM}
\begin{equation}
\begin{split}
\Sigma m \dot{r} & = \pm \sqrt{\mathcal{R}(r)}, \\
\Sigma m \dot{\theta} & = \pm \sqrt{\Theta(\theta)}.
\end{split}
\end{equation}
We can find the cyclic $t$ and $\phi$ coordinates' equation through 
\begin{equation}
\begin{split}
p_t & = \frac{\partial \mathcal{L}}{\partial \dot{t}} = \frac{\partial\left(\frac{1}{2}g_{tt}\dot{t}\dot{t}+g_{t\phi}\dot{t}\dot{\phi}\right)}{\partial \dot{t}} \\
- E & = (-1+\frac{2Mr}{\Sigma})\dot{t}-\frac{2Mra\sin^2\theta}{\Sigma} \dot{\phi}, \\
p_\phi & = \frac{\partial \mathcal{L}}{\partial \dot{\phi}} = \frac{\partial\left(g_{\phi t}\dot{\phi}\dot{t}+\frac{1}{2}g_{\phi \phi}\dot{\phi}\dot{\phi}\right)}{\partial \dot{\phi}} \\
L & = - \frac{2Mra\sin^2\theta}{\Sigma} \dot{t} + (r^2 + a^2 + \frac{2Mra^2 \sin^2\theta}{\Sigma})\sin^2\theta \ \dot{\phi}.
\end{split}
\end{equation}
Decoupling these equations gives us the equations of motion.
\begin{equation}
\begin{split}
\Sigma m \dot{\phi} & = - (aE - \frac{L}{\sin^2 \theta}) + \frac{a}{\Delta}(E(r^2+a^2)-La), \\
\Sigma m \dot{t} & = -a \sin^2 \theta (aE - \frac{L}{\sin^2 \theta}) + \frac{r^2+a^2}{\Delta}(E(r^2+a^2)-La).
\end{split}
\end{equation}
The dynamical equations can be affinely reparametrized as $m \frac{d}{d\tau} =\frac{d}{d\lambda} = \ '$ to account for the massless cases (It is now well defined for $m \rightarrow 0$ \cite{atamurotov_adujabbarov_ahmedov_2013}).
\begin{equation}
\begin{split}
\Sigma r' & = \pm \sqrt{\mathcal{R}(r)} \qquad \text{with }  \mathcal R (r) \equiv \big( (r^2+ a^2)E - aL\big)^2 - \Delta \big(m^2 r^2 + \mathcal{C}\big), \\
\Sigma \theta ' & = \pm \sqrt{\Theta(\theta)} \qquad \text{with } \Theta(\theta) \equiv \mathcal C -  \big( a\sin \theta E - \frac{L}{\sin \theta}\big)^2 - m^2 a^2 \cos^2\theta, \\
\Sigma \phi ' & = - (aE - \frac{L}{\sin^2 \theta}) + \frac{a}{\Delta}(E(r^2+a^2)-La), \\
\Sigma t' & = -a \sin^2 \theta (aE - \frac{L}{\sin^2 \theta}) + \frac{r^2+a^2}{\Delta}(E(r^2+a^2)-La).
\end{split}
\end{equation}
The sign assigned to the radial and zenithal equations depend on the direction of the motion (Radially incoming case $-\sqrt{\mathcal R(r)}$ in contrast to the radially outgoing case $+\sqrt{\mathcal R(r)}$ or the zenith value increasing $+\sqrt{\Theta (\theta)}$ or zenith value decreasing $-\sqrt{\Theta(\theta)}$).

\section{Light's radial equation and constant radial coordinate orbits}
General equations of motion for light can be written by taking the limit  $m \rightarrow 0$. Defining dimensionless quantities $\mathcal E = \frac{E}{M}$, $\mathcal L = \frac{L}{E M}$, $\Sigma^{\circ} = \frac{\Sigma}{E M}$, $\Delta^\circ = \frac{\Delta}{M^2}$, $\mathcal C^\circ = \frac{\mathcal C}{E^2M^2}$ and the new dimensionless coordinates $\rho = \frac{r}{M}$ and $T = \frac{t}{M}$.
\begin{equation}
\begin{split}
\label{eom}
\Sigma^\circ \rho' & = \pm \sqrt{\overline{\mathcal{R}} (\rho)} \qquad \text{with }  \overline{\mathcal R} (\rho) \equiv \big( \rho^2+ \alpha^2 - \alpha \mathcal L \big)^2 - \Delta^{\circ} \mathcal{C}^{\circ}, \\
\Sigma^\circ \theta ' & = \pm \sqrt{\Theta(\theta)} \qquad \text{with } \Theta(\theta) \equiv \mathcal C^\circ -   \sin^2 \theta \big( \alpha - \frac{\mathcal L}{\sin^2 \theta}\big)^2, \\
\Sigma^\circ \phi ' & = - (\alpha - \frac{\mathcal L}{\sin^2 \theta}) + \frac{\alpha}{\Delta^\circ}(\rho^2+\alpha^2- \alpha \mathcal L), \\
\Sigma^\circ T' & = - \alpha \sin^2 \theta ( \alpha - \frac{\mathcal L}{\sin^2 \theta}) + \frac{\rho^2+\alpha^2}{\Delta^\circ}(\rho^2+ \alpha^2 - \alpha \mathcal L).
\end{split}
\end{equation}
We call the coordinate $\rho_i$ at which $\rho'|_{\rho=\rho_i}=0$ \textit{radial turning points}, since that is where the particle's trajectory changes direction in the radial direction. We similarly define the turning points for other coordinates ($\theta'|_{\theta=\theta_i}=0$ and $\phi'|_{\phi=\phi_i}=0$).\newline
To find the nature of the possible orbits for our massless particle, we need to investigate the roots of $\overline{\mathcal R}=0$. Any fourth degree polynomial equation with $\rho^4$'s coefficient taken as unity can be factorized into its roots as follows
\begin{equation}
\begin{split}
\label{Rroots}
\overline{\mathcal R} & = (\rho-\rho_1)(\rho-\rho_2)(\rho-\rho_3)(\rho-\rho_4)= 0\\
& = \rho^4 - (\rho_1 + \rho_2 + \rho_3+\rho_4) \rho^3 + (\rho_1 \rho_2 + \rho_1 \rho_3 + \rho_1 \rho_4+ \rho_2 \rho_3 + \rho_2 \rho_4 + \rho_3 \rho_4) \rho^2 \\
& \qquad \qquad \qquad \qquad \qquad \qquad - (\rho_1 \rho_2 \rho_3 + \rho_1 \rho_2 \rho_4 + \rho_1 \rho_3 \rho_4 + \rho_2 \rho_3 \rho_4) \rho + \rho_1 \rho_2 \rho_3 \rho_ 4. 
\end{split}
\end{equation}
To estimate how many real roots this function \textit{could} have outside the horizon, we look into the case where all roots are real. We note that within our current case, $\overline{\mathcal R}$ has no $\rho^3$ term. This implies the that the radial equation's roots must satisfy : $\rho_1+\rho_2+\rho_3+\rho_4= 0$. One of the roots, say, $\rho_1$, has to be negative valued. To see whether we have a roots outside the event horizon $\rho_+$ ($\Delta^{\circ}|_{\rho=\rho_+} = 0$), we need to evaluate the sign of $\overline{\mathcal R}$ on it.
\begin{equation}
\begin{split}
\overline{\mathcal R}|_{\rho_+} & = \big( \rho_+^2+ \alpha^2 - \alpha \mathcal L\big)^2 = \big(2 \rho_+ - \alpha \mathcal L\big)^2 \geq 0.
\end{split}
\end{equation}
Knowing on top of this result that as $\lim_{\rho \rightarrow \infty} \overline{\mathcal{R}} \rightarrow \infty$, we surmise that outside the event horizon, we may either have two distinct real roots, two coinciding roots, or no roots at all. You can see the illustration for the sign of $\overline{\mathcal R}$ outside the event horizon for all three cases below. A physical motion is described where this function is positive valued (so long as that the square root remains real).
\begin{enumerate}
\item When there are two distinct roots outside the horizon
\begin{center}
\begin{tikzpicture}
\tkzTabInit[lgt=2,espcl=2,deltacl=0]
  {$\rho$ /.8, $\overline{\mathcal R}$ /.8}
  {,$\rho_+$,$\rho_3$,$\rho_4$,}
\tkzTabLine {,h,t,+,z,-,z,+}
\end{tikzpicture}
\end{center}
This case encompasses
\begin{enumerate}
\item Photon leaving the horizon $\rho_+$, propagating  to the turning point $\rho_3$, ending up captured by the black hole.
\item Photon coming from infinity, propagating to the turning point $\rho_4$, ending up propagating back to infinity.
\end{enumerate}
\item When there are two coinciding roots outside the horizon
\begin{center}
\begin{tikzpicture}
\tkzTabInit[lgt=2,espcl=2,deltacl=0]
  {$\rho$ /.8, $\overline{\mathcal R}$ /.8}
  {,$\rho_+$,$\rho_{3/4}$,}
\tkzTabLine {,h,t,+,z,+}
\end{tikzpicture}
\end{center}
This case encompasses
\begin{enumerate}
\item Photon leaving the horizon $\rho_+$, propagating  to the turning point $\rho_{3/4}$, where it ends up in a constant $\rho$ orbit.
\item Photon coming from infinity, propagating to the turning point $\rho_{3/4}$, where it ends up in the same constant $\rho$ orbit.
\end{enumerate}
\item There are no roots outside the event horizon
\begin{center}
\begin{tikzpicture}
\tkzTabInit[lgt=2,espcl=2,deltacl=0]
  {$\rho$ /.8, $\overline{\mathcal R}$ /.8}
  {,$\rho_+$,}
\tkzTabLine {,h,t,+}
\end{tikzpicture}
\end{center}
This case encompasses
\begin{enumerate}
\item Photon leaving the horizon $\rho_+$, propagating  to infinity
\item Photon coming from infinity, propagating to the horizon $\rho_+$
\end{enumerate}
\end{enumerate}
One should note that the nature of the roots is completely determined by an initial selection of the dimensionless impact parameters $\mathcal L$ and $\mathcal C^\circ$ related to the massless particle's initial configuration. Once this is given, the particle's radial motion will strictly be one of the above cases. For a particle coming from infinity, the various outcome of its motion (falls into the black hole, orbits circularly or escapes back to infinity) are distinguished by some critical values of the dimensionless impact parameters \cite{frolov_novikov_1998}. These ideas will be further developed while studying the shadow of a black hole. \newline
Let us briefly investigate the stability of the constant $\rho$ orbits (When $\overline{\mathcal R} = 0$ and $\frac{d \overline{\mathcal R}}{dr}=0$). It's stability can be checked by seeing how $\overline{\mathcal R}$ behaves around these coordinates, which can be argued from the Taylor series expansion of $\overline{\mathcal R}$ around that point
\begin{equation}
\overline{\mathcal R} (\rho) = \sum_{n=0}^{\infty} \frac{\overline{\mathcal R}^{(n)}|_{\rho_{3/4}}}{n!}(\rho - \rho_{3/4})^n.
\end{equation} 
from which the first two terms vanish due to our roots being coinciding turning points. For a small perturbation $\xi$ around our constant radius orbit $\rho = \rho_{3/4} + \xi$, our expansion becomes
\begin{equation}
\overline{\mathcal R} (\xi) =  \frac{\overline{\mathcal R}^{(2)}|_{\rho_{3/4}}}{2} \xi^2 + \mathcal O (\xi^3).
\end{equation}
Where we can deduce from our second sign table that $\overline{\mathcal R}^{(2)}|_{\rho_{3/4}}$ is positive valued. This implies $\overline{\mathcal R} (\xi)$ will be real valued, which allows radial perturbation to be carried out without restrain. This behaviour makes the orbit unstable.

\section{Constants of motion and restrictions for a general massless case}
Manipulating the general equations for a massless case back in \eqref{eom}, specifically defining Carter's constant $\mathcal Q \equiv \mathcal C - (a E - L)^2$, or in dimensionless form, $\mathcal Q^\circ \equiv \mathcal C^\circ - (\alpha - \mathcal L)^2 $, we simplify the behaviour of the massless trajectory in the zenithal and radial directions. Writing out $\overline{\mathcal R} (\rho)$ and $\Theta (\theta)$ in terms of this new quantity,
\begin{equation}
\begin{split}
\overline{\mathcal R} (\rho) & = \big( \rho^2+ \alpha^2 - \alpha \mathcal L \big)^2 - \Delta^{\circ} \mathcal{C}^{\circ} = \rho^4 + \Big( 2 \alpha (\alpha - \mathcal L) - \mathcal C^\circ \Big) \rho^2 + 2 \mathcal C^\circ \rho - \alpha^2 \Big( \mathcal C^\circ - (\alpha - \mathcal L)^2 \Big) \\
& = \rho^4 + \Big( (\alpha - \mathcal L)(\alpha + \mathcal L) - \mathcal Q ^\circ \Big) \rho^2 + 2 (\mathcal Q^\circ + (\alpha - \mathcal L)^2) \rho - \alpha^2 \mathcal Q^\circ. \\
\end{split}
\end{equation}
And
\begin{equation}
\begin{split}
\Theta(\theta) & = \mathcal C^\circ - (\alpha \sin \theta - \frac{\mathcal L}{\sin \theta})^2 = \underbrace{\mathcal C^\circ - (\alpha - \mathcal L)^2}_{\equiv \mathcal Q^\circ} + (\alpha - \mathcal L)^2 - (\alpha \sin \theta - \frac{\mathcal L}{\sin \theta})^2 \\
& = \mathcal Q^\circ + \alpha^2 \cancelto{i}{- 2 \alpha \mathcal L} + \mathcal L^2 - \alpha \sin ^2 \theta \cancelto{i}{+ 2 \alpha \mathcal L} - \frac{\mathcal L^2}{\sin ^2 \theta} = \mathcal Q^\circ + \alpha \cos ^2 \theta - \mathcal L^2 \cot ^2 \theta \\
& =  \mathcal Q^\circ + \cos ^2 \theta \Big( \alpha^2 - \frac{\mathcal L^2}{\sin ^2 \theta} \Big).
\end{split}
\end{equation}
We can write the equation of motion as
\begin{equation}
\begin{split}
\Sigma^\circ \rho' & = \pm \sqrt{\overline{\mathcal{R}} (\rho)} \qquad \text{with }  \overline{\mathcal R} (\rho) \equiv  \rho^4 + \Big( \alpha^2 -\mathcal L^2 - \mathcal Q ^\circ \Big) \rho^2 + 2 (\mathcal Q^\circ + (\alpha - \mathcal L)^2) \rho - \alpha^2 \mathcal Q^\circ, \\
\Sigma^\circ \theta ' & = \pm \sqrt{\Theta(\theta)} \qquad \text{with } \Theta(\theta) \equiv \mathcal Q^\circ + \cos ^2 \theta \Big( \alpha^2 - \frac{\mathcal L^2}{\sin ^2 \theta} \Big), \\
\Sigma^\circ \phi ' & = - (\alpha - \frac{\mathcal L}{\sin^2 \theta}) + \frac{\alpha}{\Delta^\circ}(\rho^2+\alpha^2- \alpha \mathcal L), \\
\Sigma^\circ t' & = - \alpha \sin^2 \theta ( \alpha - \frac{\mathcal L}{\sin^2 \theta}) + \frac{\rho^2+\alpha^2}{\Delta^\circ}(\rho^2+ \alpha^2 - \alpha \mathcal L).
\end{split}
\end{equation}
We now focus on the zenithal equation to find restrictions on the impact parameters that might exclude a certain photons irrelevant to our study. Substituting $u = \cos \theta$ and thus $u' = - \theta ' \sin \theta$, we get
\begin{equation}
\Sigma^\circ u' = \pm \sqrt{\Theta(u)} \qquad \text{with } \Theta(u) \equiv - \alpha^2 u^4 + \Big( \alpha^2 - \mathcal L^2 - \mathcal Q^\circ \Big) u^2 + \mathcal Q^\circ
\end{equation}
The motion is only possible for coordinates for which $\Theta (u) \geq 0$. The coordinates for which $\Theta (u) = 0$ are the zenithal turning points of the motion. At the extremities $u=1$ and $u=-1$ (or $\theta = 0$ and $\theta = \pi$), our function becomes negative : $\Theta (1) = \Theta (-1) = - \mathcal L ^2$, meaning that for an approaching photon, we should have at least two root $0 \leq u_\pm^2 \leq 1$ between which there can be angles for which $\Theta(u)$ can become positive. \newline
But of course, the nature of the roots will depend on the physical quantities $\alpha$, $\mathcal L$ and $\mathcal Q^\circ$
\begin{equation}
u^2_\pm = \frac{\alpha^2-\mathcal L^2 - \mathcal Q^\circ \pm \sqrt{(\alpha^2-\mathcal L^2 - \mathcal Q^\circ)^2 + 4 \alpha^2 \mathcal Q^\circ}}{2\alpha^2}
\end{equation}
Let us demonstrate how these roots change according to the sign of $\mathcal Q^\circ$
\begin{enumerate}[label=(\roman*)]
\item $\mathcal Q^\circ > 0$ \newline
The only root among $u^2_\pm$ to be positive is the $u_+ ^2$ root. This leaves us with two final roots for $u$ ; $- u_+$ and $u_+$. The sign of $\Theta$ can be  
\begin{center}
\begin{tikzpicture}
\tkzTabInit[lgt=2,espcl=2,deltacl=0]
  {$u$ /.8, $\Theta$ /.8}
  {,$-1$,$-u_+$,$u_+$,$1$,}
\tkzTabLine {,h,t,-,z,+,z,-,t,h,}
\end{tikzpicture}
\end{center}
Which simply corresponds to the particle going back and forth between the angles corresponding to $-u_+$ and $u_+$, while passing through the equatorial plane.
\item $\mathcal Q^\circ = 0$ \newline
The possible roots become $u_-^2 = 0$ and $u_+^2 = 1 - \frac{\mathcal L^2}{\alpha ^2}$. The $u_-^2$ root is always a valid one, while the $u_+^2$ root can only be valid for $\mathcal L \leq \alpha$. \newline
For a motion with $\mathcal L > \alpha$, the sign of $\Theta$ goes as
\begin{center}
\begin{tikzpicture}
\tkzTabInit[lgt=2,espcl=2,deltacl=0]
  {$u$ /.8, $\Theta$ /.8}
  {,$-1$,$\pm u_-$,$1$,}
\tkzTabLine {,h,t,-,z,-,t,h,}
\end{tikzpicture}
\end{center}
which corresponds to a motion constrained to the equatorial plane. \newline
While for $\mathcal L \leq \alpha$, the sign of $\Theta$ goes as
\begin{center}
\begin{tikzpicture}
\tkzTabInit[lgt=2,espcl=2,deltacl=0]
  {$u$ /.8, $\Theta$ /.8}
  {,$-1$,$-u_+$,$\pm u_-$,$u_+$,$1$,}
\tkzTabLine {,h,t,-,z,+,z,+,z,-,t,h,}
\end{tikzpicture}
\end{center}
which, in turn, corresponds to a motion which can oscillate between the angles corresponding to $- u_+$ and $u_+$, with an unstable equatorial angle motion.
\item $\mathcal Q^\circ < 0$ \newline
The sum of the roots can be expressed as $u_-^2 + u_+^2 = \frac{\alpha^2 - \mathcal L ^2 - \mathcal Q^\circ}{\alpha^2}$, while their product gives $u_-^2 u_+^2 = \frac{- \mathcal Q^\circ}{\alpha^2} > 0$. This implies the roots are either both positive or both negative at the same time (none of them are zero). In order to have both of them positive valued, we require $\alpha^2 - \mathcal L ^2 - \mathcal Q^\circ > 0$ to strictly hold. We shall see why in our case, this is ruled out physically.
\end{enumerate}
The analysis above is valid for any general motion. \newline
For our \textit{black hole shadow} application, we want to understand the behaviour of photons coming from very far away from the black hole ("very far" is used interchangeably with "infinity" in this context). Which ones among these photons make it to an observer very far away, and which ones are captured by our black hole ? This question is readily answered by the radial equation on its own. As seen previously in Section 5, depending on the impact parameters $\mathcal L$ and $\mathcal Q^\circ$, the particle coming from infinity may encounter a turning point $\rho_4$ before the horizon (or it may not). If it does, this means the photon will bounce back from this point back to infinity, meaning it can reach our observer. Since the function $\overline{\mathcal R}$ is continuous in these parameters, the roots $\rho_{3/4}$ will also show similar continuous behaviour in these parameters. This indicates a smooth passage from the case of two distinct roots outside the event horizon (Scenario 1 in Section 5) to the case of two coinciding roots (Scenario 2 in Section 5), and from there, to the case of no roots (Scenario 3 in Section 5) as we continuously vary the impact parameters. One can infer from this that the case of coinciding roots -unstable constant $\rho$ orbits- is effectively separating a captured photon from an escaping one. We shall now inspect these critical impact parameters $\mathcal L_c$ and $\mathcal Q^\circ_c$ in the ($\mathcal L$,$\mathcal Q^\circ$) parameter space. We will notice that we can represent these with a parametric curve. \newline
To this end, we now focus on what impact parameters are required for our photon to be entering constant $\rho$ orbits. We require $\overline{\mathcal R}(\rho_c) = \frac{d \overline{\mathcal R}}{d \rho}|_{\rho = \rho_c} = 0$ \cite{teo_2003}, where $\rho_c$ denotes the dimensonless radial coordinate at which a photon coming in with some specific parameters $\mathcal L$ and $\mathcal Q^\circ$ exhibits circular orbit. Solving the system of equations for these parameters gives us two class of solutions in terms of the constant $\rho_c$ of the orbit:
\begin{enumerate}[label=(\alph*)]
\item $\mathcal L_a = \frac{\rho_c^2 + \alpha^2}{\alpha}$, $\mathcal Q^\circ_a = - \frac{\rho_c^4}{\alpha^2}$. \newline
\item $\mathcal L_b = \frac{\alpha(\rho_c + 1) + \rho_c^2 (\rho_c - 3)}{\alpha (1-\rho_c)}$, $\mathcal Q^\circ_b = \frac{\rho_c^3\big(4\alpha^2 - \rho_c(\rho_c-3)^2\big)}{\alpha^2(1-\rho_c)^2}$.
\end{enumerate}
Among (a) class solutions, we can quickly see how $\mathcal Q^\circ_a < 0$ is always the case. We saw earlier in our analysis that in order to have some valid roots $u_\pm^2$ of the $\Theta$ function, we require $\alpha^2 - \mathcal L_a ^2 - \mathcal Q^\circ_a > 0$ to hold. Writing the inequality out explicitly for (a) ;
\begin{equation}
\alpha^2 - \mathcal L_a ^2 - \mathcal Q^\circ_a = - 2 \rho^2.
\end{equation}
Which is strictly negative valued for real $\rho$. This implies that this first class of solution cannot describe a motion which eventually ends up in a constant $\rho$ orbit, and so, is no use to us. \newline
We now turn our attention to the solution class (b). $\mathcal Q^\circ _b$ has no immediate sign that can be assigned to it, but let us check the sign of $\alpha^2 - \mathcal L_a ^2 - \mathcal Q^\circ_a > 0$ to see if we can exclude some region of $\mathcal Q^\circ _b$. Writing the inequality out explicitly for (b)
\begin{equation}
\alpha^2 - \mathcal L_b ^2 - \mathcal Q^\circ_b = \frac{- 2\rho_c\big( \rho_c (\rho_c^2-3) + 2 \alpha^2 \big)}{(1-\rho_c)^2}.
\end{equation}
The sign of this function is not clear at first sight. The drawn out examples below illustrates the sign of this quantity for any $\rho_c$.

\begin{figure}[H]
\centering
\subfloat[][$\alpha = 0$]{
  \includegraphics[width=0.3\columnwidth]{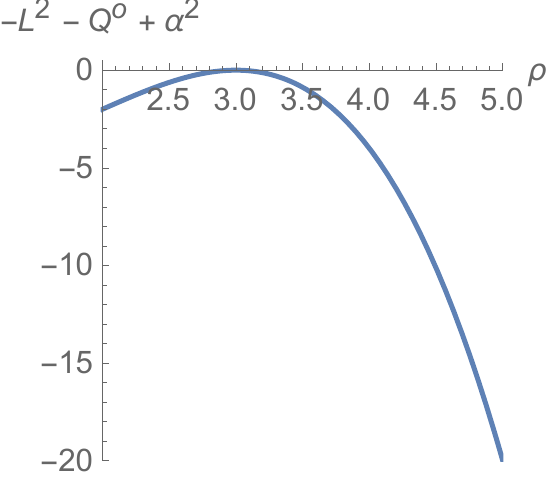}
}\hspace{3cm}
\subfloat[][$\alpha = 0.33$]{
  \includegraphics[width=0.3\columnwidth]{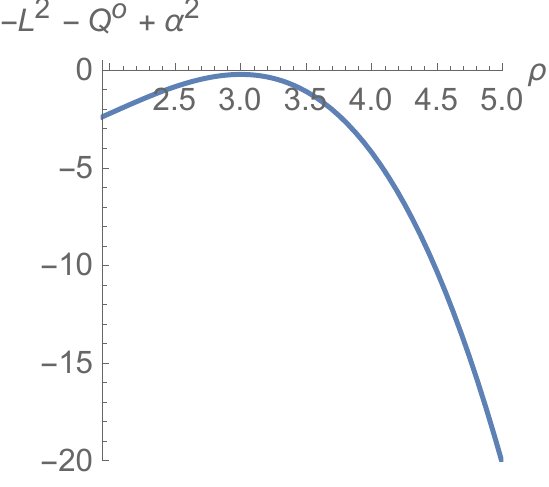}
}
\end{figure}
\begin{figure}[H]
\centering
\subfloat[][$\alpha = 0.67$]{
  \includegraphics[width=0.3\columnwidth]{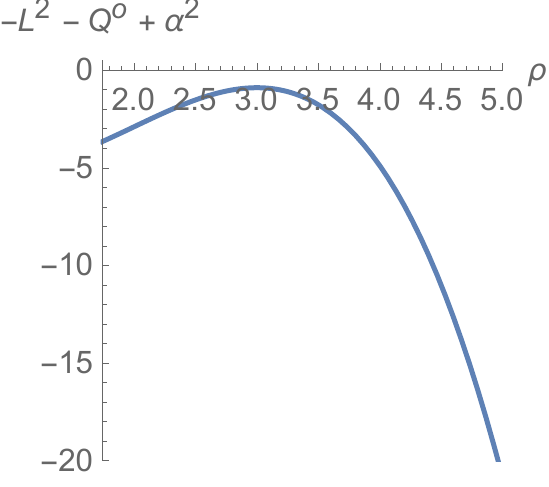}
}\hspace{3cm}
\subfloat[][$\alpha = 1$]{
  \includegraphics[width=0.3\columnwidth]{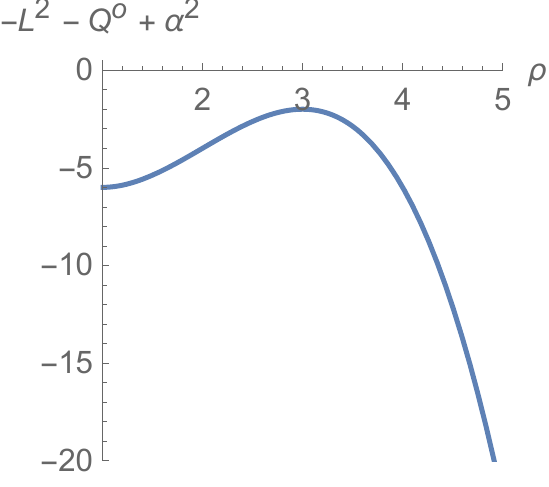}
}

\caption{The value graphs of $\alpha^2 - \mathcal L_b ^2 - \mathcal Q^\circ_b$ as we scan through different $\rho_c$ values for different $\alpha$. Its values are negative.}

\end{figure}

We thus surmise that this class of solutions also cannot admit $\mathcal Q^\circ < 0$ as a motion ending up in a constant $\rho$ orbit. The orbits of interest are thus strictly satisfying $\mathcal Q^\circ_b \geq 0$, and possess class (b) type of impact parameters only. For notation's sake, we shall get rid of the $b$ subscripts and denote these critical parameters as $\mathcal L_c$ and $\mathcal Q^\circ _c$. \newline
We can imagine $\mathcal L_c$ and  $\mathcal Q^\circ_c$ as parametric curves in the ($\mathcal L$, $\mathcal Q^\circ$) parameter space, defined through $\rho_c$. As we "scan" through values of $\rho_c$ (equivalent to ``scanning'' for photons coming from different polar angles in the observer's sky), we are essentially finding the points on the parameter space for which a constant $\rho=\rho_c$ motion happens.

\begin{figure}[H]
\centering
\subfloat[][]{
  \includegraphics[width=0.6\columnwidth]{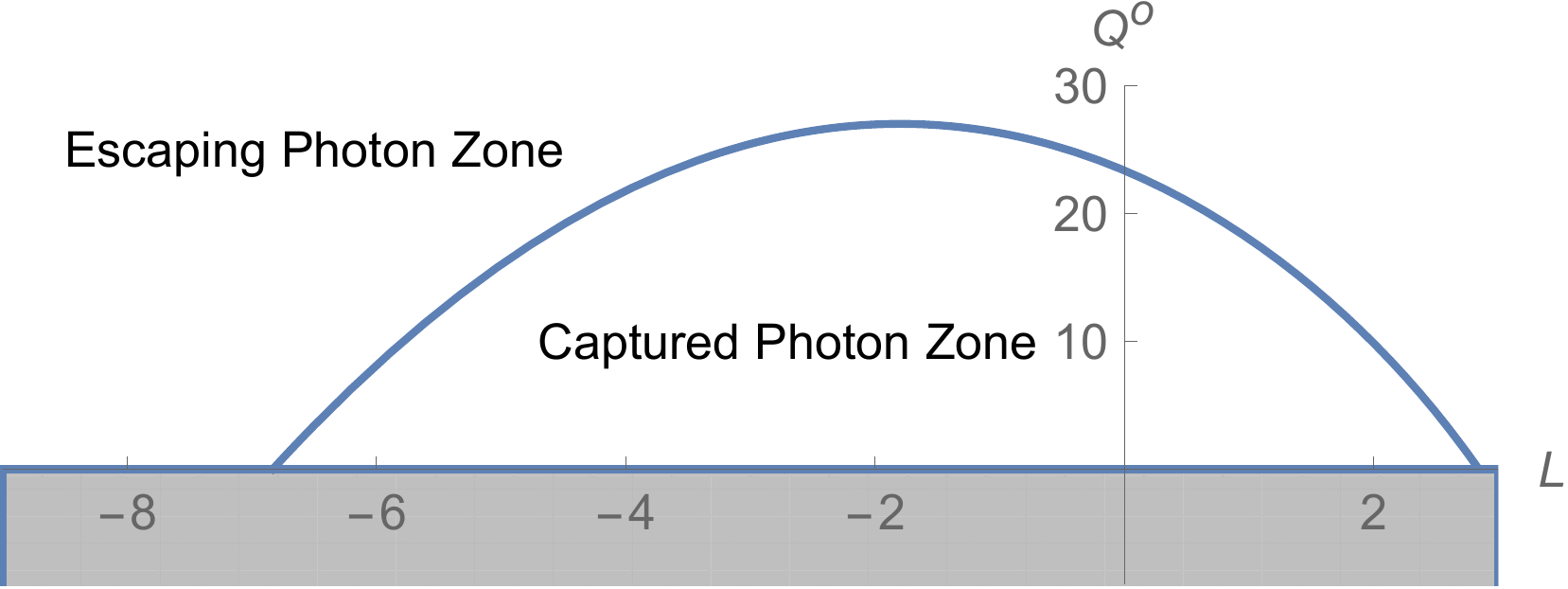}
}
\caption{Drawn out parameter space ($\mathcal L$, $\mathcal Q^\circ$) for $\alpha = 0.9$, where the curve drawn are the critical parameters $\mathcal L_c$ and  $\mathcal Q^\circ_c$, separating the parameters with which a photon can escape -\textit{Escaping Photon Zone}- and with which a photon is captured -\textit{Captured photon zone}-. The values for which $\mathcal Q^\circ < 0$ cannot admit critical photons.}
\end{figure}

All photons with parameters inside the \textit{Captured Photon Zone} are eventually captured by the black hole. Any photon with parameters on top of the curve are eventual constant $\rho$ orbit photons. Photons with any other parameter finally escapes the black hole. \cite{vazquez_esteban_2018} \newline
The captured photons are evidently invisible to someone looking at the black hole. The question motivating the next chapter is then: ``What regions of the sky stays in the \textit{shadow} of this black hole for an observer looking at it from a distance ?''.

\section{Observer's sky and Black hole Shadow}
Assume the space in question only contains this Kerr black hole, an observer and a source of light. Both the observer and the source of light are taken to be very far away from the black hole, enough so that the observer is seeing its neighborhood as flat spacetime (which is due to this space being taken as asymptotically flat). This observer can thus pick a cartesian space coordinate system with its origin taken at the center of the black hole, while picking the orientation such that its positioned at $\vec{r}_0$ with no azimuthal components, while having some zenithal angle $\theta_0$. Of course, around the black hole, where the space isn't flat, this coordinate system does not coincide with the Boyer-Lindquist coordinates introduced above. Only at infinities does the $x^2+y^2+z^2 = r^2$ relation hold, and the spaces agree on distances between points \cite{vazquez_esteban_2018}. As far as the observer is concerned, photons arrive to it in this locally flat spacetime, in some direction in its perceived sky. This \textit{Observer's Sky} is essentially a 2 dimensional dome, but since we are only interested in the incoming light from around the black hole (which represents a very small solid angle of this dome), we can approximate this part of the sky to some 2 dimensional plane, whose embedding is illustrated in the figure below. The plane's points can be described with some 2 dimensional cartesian coordinate $(\xi,\eta)$.

\begin{figure}[H]
\centering
\subfloat[][]{
  \includegraphics[width=0.6\columnwidth]{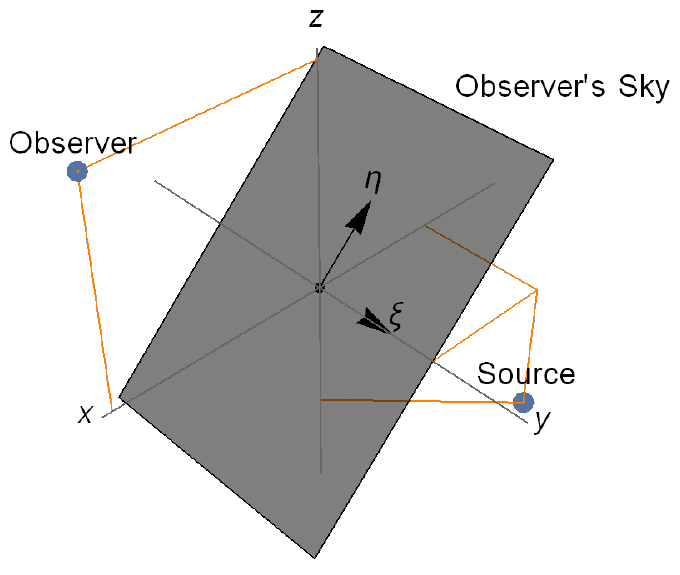}
}
\caption{Illustration of the observer's sky coordinate system as embedded into the 3-dimensional space. Since the neighborhood of the black hole represents a very small portion of the observer's sky, we approximate this part of the sky as a plane.}
\end{figure}

We should note here that a point $(\xi,\eta)$ in the observer's sky (also called the celestial plane) is embeded onto $(-\eta \cos \theta_0 , \xi , \eta \sin \theta_0 )$ in the observer's coordinate system. \newline
The incoming photon's trajectory curve in this 3 dimensional space can be expressed parametrically in terms of its distance to the black hole (which is monotonically increasing around the observer) as follows
\begin{equation}
\vec{r}_\gamma = \big( x(r), y(r), z(r) \big).
\end{equation}
Whose tangent vector at the observer's position $\vec{r}_0$ can be expressed simply as
\begin{equation}
\vec{v}_\gamma|_{\vec{r}_0} = \big( \frac{dx}{dr}|_{\vec{r}_0}, \frac{dy}{dr}|_{\vec{r}_0}, \frac{dz}{dr}|_{\vec{r}_0} \big).
\end{equation}
Tracing back this vector from the observer's position to the celestial plane,
\begin{equation}
\vec{r}_0 - r_0 \vec{v}_\gamma|_{\vec{r}_0} = (r_0 \sin \theta_0 - r_0  \frac{dx}{dr}|_{\vec{r}_0}, - r_0 \frac{dy}{dr}|_{\vec{r}_0}, r_0 \cos \theta_0 - r_0 \frac{dz}{dr}|_{\vec{r}_0}).
\end{equation}
Where we have taken the negative of the tangent vector to trace the photon backwards, and we have used $r_0$ as a multiplier since all of the points around the black hole and on top of this celestial plane are approximately $r_0$ distance away from the observer.

\begin{figure}[H]
\centering
\subfloat[][]{
  \includegraphics[width=0.6\columnwidth]{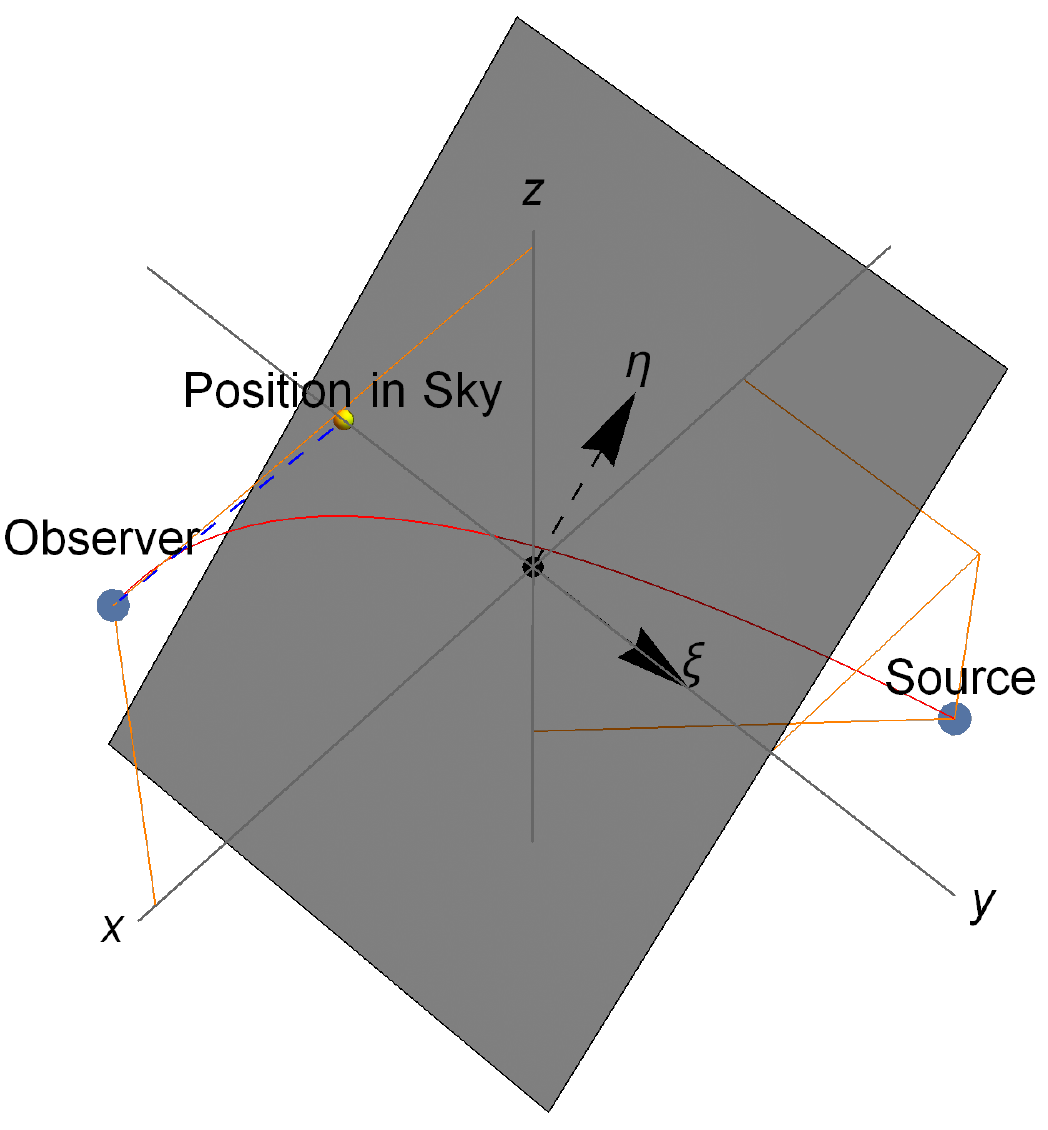}
}
\caption{Illustration of the incoming photon's real trajectory (red), and the tracing back of the incident photon onto the celestial plane (dashed, blue).}
\end{figure}

We can now relate the celestial coordinates $\xi$ and $\eta$ to these quantities by using the embedding mentionned in the previous page. We simply equate each of the $x$, $y$ and $z$ components individually :
\begin{equation}
\begin{split}
\label{planareqns}
-\eta \cos \theta_0 & =  r_0 \sin \theta_0 - r_0  \frac{dx}{dr}|_{\vec{r}_0} \qquad \qquad \, x \text{ component}, \\
\xi & =  - r_0 \frac{dy}{dr}|_{\vec{r}_0} \qquad \qquad \qquad \qquad y \text{ component}, \\
\eta \sin \theta_0 & =  r_0 \cos \theta_0 - r_0 \frac{dz}{dr}|_{\vec{r}_0} \qquad \qquad \, z \text{ component}.
\end{split}
\end{equation}
Going to spherical coordinates using $x=r \sin \theta \cos \phi$, $y=r \sin \theta \sin \phi$ and $z=r \cos \theta$, which in turn implies
\begin{equation}
\begin{split}
\frac{dx}{dr} & = \sin \theta \cos \phi + r \cos \theta \cos \phi \frac{d \theta}{d r} - r \sin \theta \sin \phi \frac{d \phi}{d r} \\
\frac{dx}{dr}|_{\vec{r}_0} & = \sin \theta_0 + r_0 \cos \theta_0 \frac{d \theta}{d r}|_{\vec{r}_0}, \\
\frac{dy}{dr} & = \sin \theta \sin \phi + r \cos \theta \sin \phi \frac{d \theta}{d r} + r \sin \theta \cos \phi \frac{d \phi}{dr} \\
\frac{dy}{dr}|_{\vec{r}_0} & = r_0 \sin \theta_0 \frac{d \phi}{dr}|_{\vec{r}_0}, \\
\frac{dz}{dr} & = \cos \theta - r \sin \theta \frac{d \theta}{dr} \\
\frac{dz}{dr}|_{\vec{r}_0} & = \cos \theta_0 - r_0 \sin \theta_0 \frac{d \theta}{dr}|_{\vec{r}_0}.
\end{split}
\end{equation}
We can now solve \eqref{planareqns} for $\xi$ and $\eta$
\begin{equation}
\begin{split}
\label{celcoord1}
- \eta \cos \theta_0 & =\cancelto{}{r_0 \sin \theta_0} - \cancelto{}{r_0 \sin \theta_0} - r_0^2 \cos \theta_0 \frac{d\theta}{dr}|_{\vec{r}_0} \\
\eta \ \cancelto{}{\cos \theta_0} & = r_0^2 \ \cancelto{}{\cos \theta_0} \ \frac{d\theta}{dr}|_{\vec{r}_0} \qquad \qquad \text{Since } \cos \theta_0 \neq 0 \text{ in general.} \\
\eta & = r_0^2 \frac{d\theta}{dr}|_{\vec{r}_0}.
\end{split}
\end{equation}
It can be checked similarly that the third equation of \eqref{planareqns} gives the same relation. The result for the second equation is straightforward
\begin{equation}
\begin{split}
\label{celcoord2}
\xi = - r_0^2 \sin \theta_0 \frac{d \phi}{dr}|_{\vec{r}_0}.
\end{split}
\end{equation}
For large $r_0$, we can approximate our $\frac{d\theta}{dr}|_{\vec{r}_0}$ and $\frac{d \phi}{dr}|_{\vec{r}_0}$ terms using our dynamical equations.
\begin{equation}
\begin{split}
\frac{d\theta}{d \rho} & = \pm \sqrt{\frac{\mathcal Q^\circ + \cos ^2 \theta (\alpha^2 - \frac{\mathcal L^2}{\sin ^2 \theta})}{\rho^4 + \Big( \alpha^2 -\mathcal L^2 - \mathcal Q ^\circ \Big) \rho^2 + 2 (\mathcal Q^\circ + (\alpha - \mathcal L)^2) \rho - \alpha^2 \mathcal Q^\circ}} \\
\frac{d\theta}{d \rho}|_{\vec{r}_0} & = \pm \frac{1}{\rho_0^2} \sqrt{\mathcal Q^\circ + \cos ^2 \theta_0 (\alpha^2 - \frac{\mathcal L^2}{\sin ^2 \theta_0})} \\
\frac{d\theta}{d r}|_{\vec{r}_0} & = \pm \frac{M}{r_0^2}\sqrt{\mathcal Q^\circ + \cos ^2 \theta_0 (\alpha^2 - \frac{\mathcal L^2}{\sin ^2 \theta_0})}.
\end{split}
\end{equation}
And
\begin{equation}
\begin{split}
\frac{d\phi}{d\rho} & = \frac{-(\alpha - \frac{\mathcal L}{\sin^2 \theta}) + \frac{\alpha}{\Delta^\circ}(\rho^2 + \alpha^2 - \alpha \mathcal L)}{\sqrt{\rho^4 + \Big( \alpha^2 -\mathcal L^2 - \mathcal Q ^\circ \Big) \rho^2 + 2 (\mathcal Q^\circ + (\alpha - \mathcal L)^2) \rho - \alpha^2 \mathcal Q^\circ}} \\
\frac{d\phi}{d\rho}|_{\vec{r}_0} & = \frac{1}{\rho_0^2}\frac{\mathcal L}{\sin^2 \theta_0} \\
\frac{d\phi}{dr}|_{\vec{r}_0} & = \frac{M}{r_0^2} \frac{\mathcal L}{\sin^2 \theta_0}.
\end{split}
\end{equation}
Where in both results, we have used the fact that the radial equation's sign (the "value" of the $\pm$) is taken to be positive for a radially outgoing photon, while the sign of the zenithal equation can be both positive or negative, depending on the motion. \newline
We can thus finally write out the explicit relations for the celestial coordinates
\begin{equation}
\begin{split}
\label{coordrel}
\xi & = - r_0^2 \sin \theta_0 \frac{M}{r_0^2} \frac{\mathcal L}{\sin^2 \theta_0} \\
& = - M \frac{\mathcal L}{\sin \theta_0}, \\
\eta & = \pm r_0^2 \frac{M}{r_0^2} \sqrt{Q^\circ + \cos^2 \theta_0 (\alpha^2-\frac{\mathcal L^2}{\sin^2 \theta_0})} \\
& = \pm M \sqrt{Q^\circ + \cos^2 \theta_0 (\alpha^2 - \frac{\mathcal L^2}{\sin^2 \theta_0})}.
\end{split}
\end{equation}
One can check that these coordinates satisfy the following equation (Defining new dimensionless celestial coordinates $\xi^\circ = \frac{\xi}{M}$ and $\eta^\circ = \frac{\eta}{M}$)
\begin{equation}
\label{skyshadow}
(\xi^\circ - \alpha \sin \theta_0)^2 + \eta^{\circ \ 2} = \mathcal Q^\circ + (\mathcal L+ \alpha)^2.
\end{equation}
What this equation tells us is that while we look around a Kerr black hole with spin parameter $\alpha$, photons coming from celestial coordinates ($\xi^\circ$,$\eta^\circ$) possesses impact parameters $\mathcal Q^\circ$ and $\mathcal L$ satisfying the above equation. \newline
One form of \eqref{skyshadow} that will be usefull to us soon is writing it out in terms of the critical radial coordinate $\rho_c$, which we can do by substituting the class (b) solutions of $\mathcal Q^\circ_c$ and $\mathcal L_c$ into it
\begin{equation}
\label{shadowoutline}
(\xi^\circ-\alpha \sin \theta_0)^2 + \eta^{\circ \ 2} = \frac{4(\alpha^2 + \rho_c^2 (2\rho_c - 3))}{(\rho_c-1)^2}.
\end{equation}
We could note here that the extermities of the black hole shadow have the same form as a parametric polar plot, where the parameter expressed in the last expression is the $\rho_c$ parameter for a fixed black hole of spin parameter $\alpha$.

\begin{figure}[H]
\centering
\subfloat[][]{
  \includegraphics[width=0.4\columnwidth]{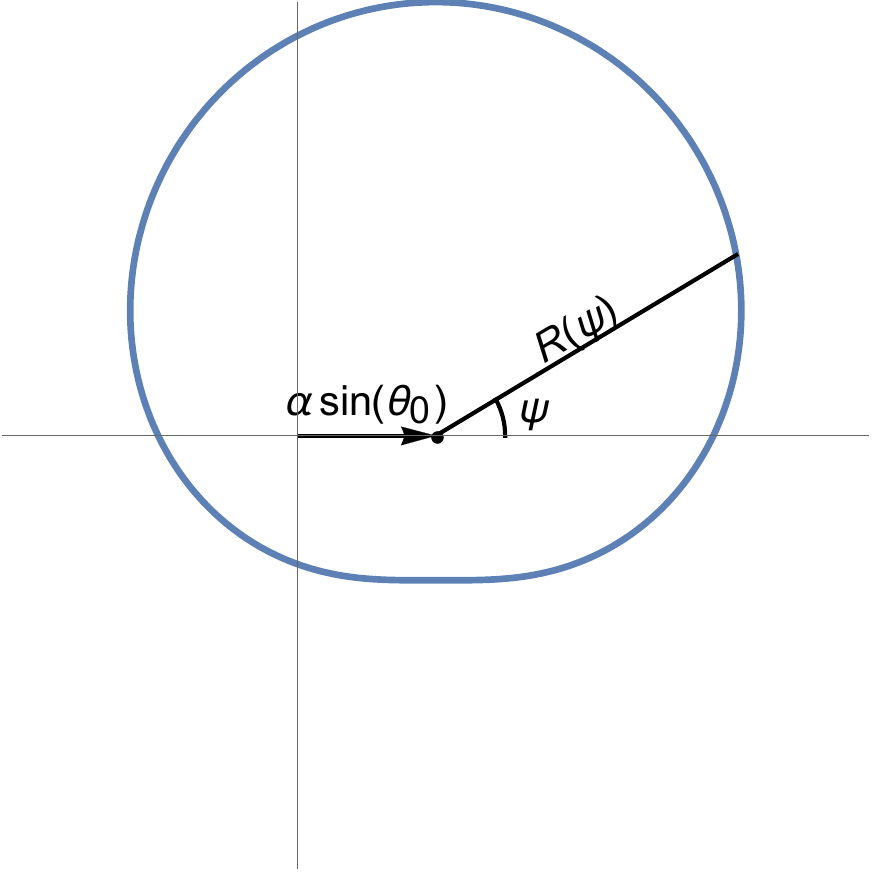}
}
\caption{An example of a parametric polar plot, with parameter $\psi$, which corresponds to the polar angle on this graph. The illustration is that of the curve \newline $(\xi^\circ-\alpha \sin \theta_0)^2 + \eta^{\circ \ 2} = R(\psi)^2$, where $R(\psi) = 2 + \sin(\psi)$}
\end{figure}

For each polar angle $\psi$ for the drawn out shadow, we have a corresponding critical radius $\rho_c$ for photons coming in from that polar angle in our observer's sky. 

\begin{figure}[H]
\centering
\subfloat[][]{
  \includegraphics[width=0.4\columnwidth]{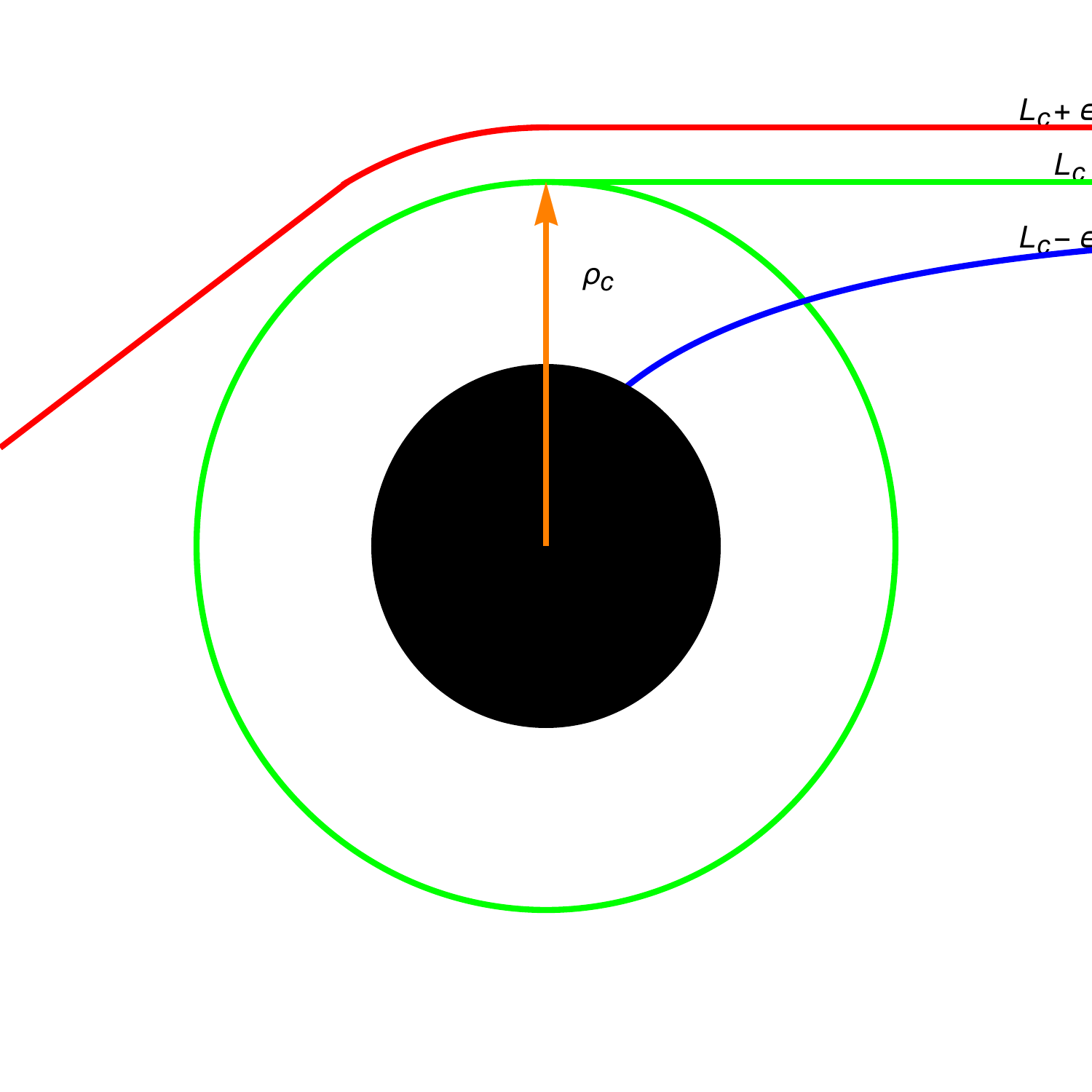}
}
\caption{A crude illustration of how slight variations around the critical impact parameter $\mathcal L_c$ changes the orbits of incoming photons -For the same impact parameter $\mathcal Q^\circ$-. The photon with red trajectory escapes to the observer. The green trajectory photon ends up in a constant radius orbit, while the blue trajectory photon ends up captured by the black hole.}
\end{figure}
 \newpage
\section{Drawn out examples of black hole shadows}
Let us consider two extreme cases and draw out the corresponding outlines of their shadows
\subsection{$\alpha \ll 1$ for an observer in the equatorial plane $\theta_0 = \pi/2$}
Let us remember the two conditions needed to be satisfied by the critical radius $\rho_c$ : $\overline{\mathcal R}(\rho_c) = 0$ and $\frac{d \overline{\mathcal R}}{d\rho}|_{\rho_c} = 0$. One simple trick we might use to facilitate our calculations would be to note that if the mentioned system of equations are correct, so are $\frac{\overline{\mathcal R}(\rho_c)}{\rho_c} = 0$ and $\frac{d}{d\rho}(\frac{\overline{\mathcal R}}{\rho})|_{\rho_c} = 0$. These correspond to the following equations, respectively
\begin{equation}
\begin{split}
\rho_c^3 + (\mathcal L_c^2 + \mathcal Q_c^\circ) \rho_c + 2 (\mathcal L_c^2 + \mathcal Q_c^\circ - 2 \alpha \mathcal L_c) + \mathcal O (\alpha^2)& = 0 \\
3 \rho_c^2 - (\mathcal L_c^2 + \mathcal Q_c^\circ) +  \mathcal O (\alpha^2) &= 0.
\end{split}
\end{equation}
Using the second equation to substitute $\mathcal Q_c^\circ$ into the first one, we get
\begin{equation}
-2\rho_c \big( \rho_c^2(\rho_c-3)+2\alpha \mathcal L_c \big) = 0.
\end{equation}
Solving this perturbatively (by substituting $\rho_c = \rho_0 + \alpha \rho_1$), then equating $\alpha$ terms to one another, one finds
\begin{equation}
\rho_c = 3 - \alpha \frac{2 \mathcal L_c}{9} + \mathcal O (\alpha^2).
\end{equation}
Plugging this back into \eqref{shadowoutline}, we get the outlines of the shadow of the black hole as a relation on the sky coordinates
\begin{equation}
(\xi^\circ - \alpha)^2 + \eta^{\circ \ 2} = 27 - 2 \alpha \mathcal L_c.
\end{equation}
One could extract from \eqref{coordrel} that for this case ($\theta_0 = \pi/2$), $\mathcal L_c = - \xi^\circ$, thus simpliying the relation to
\begin{equation}
\begin{split}
(\xi^\circ- \alpha)^2 + \eta^{\circ \ 2} & = 27+ 2 \alpha \xi^\circ \\
(\xi^\circ- 2\alpha)^2 + \eta^{\circ \ 2} & = 27.
\end{split}
\end{equation}
The shadow has the same radius as a Schwarzschild Black Hole of the same mass ($\rho_{\text{Sch}}=3\sqrt{3}$), but the center of the shadow is shifted by $2 \alpha$ towards the positive $\xi^\circ$ direction, or in other words, "towards" the rotation of the black hole.

\begin{figure}[H]
\centering
\subfloat[][$\alpha = 0$]{
  \includegraphics[width=0.3\columnwidth]{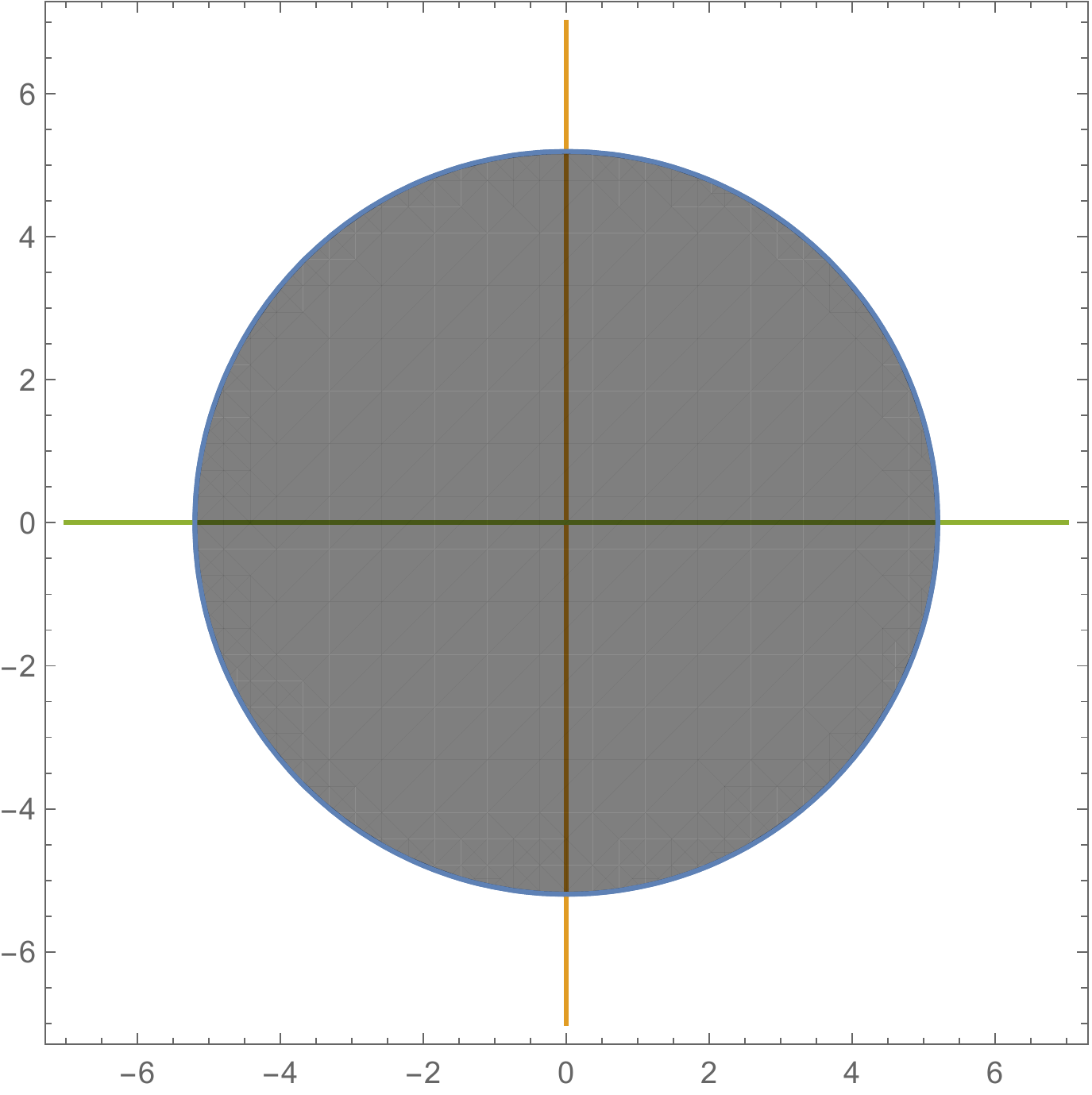}
}\hspace{3cm}
\subfloat[][$\alpha = 0.3$]{
  \includegraphics[width=0.3\columnwidth]{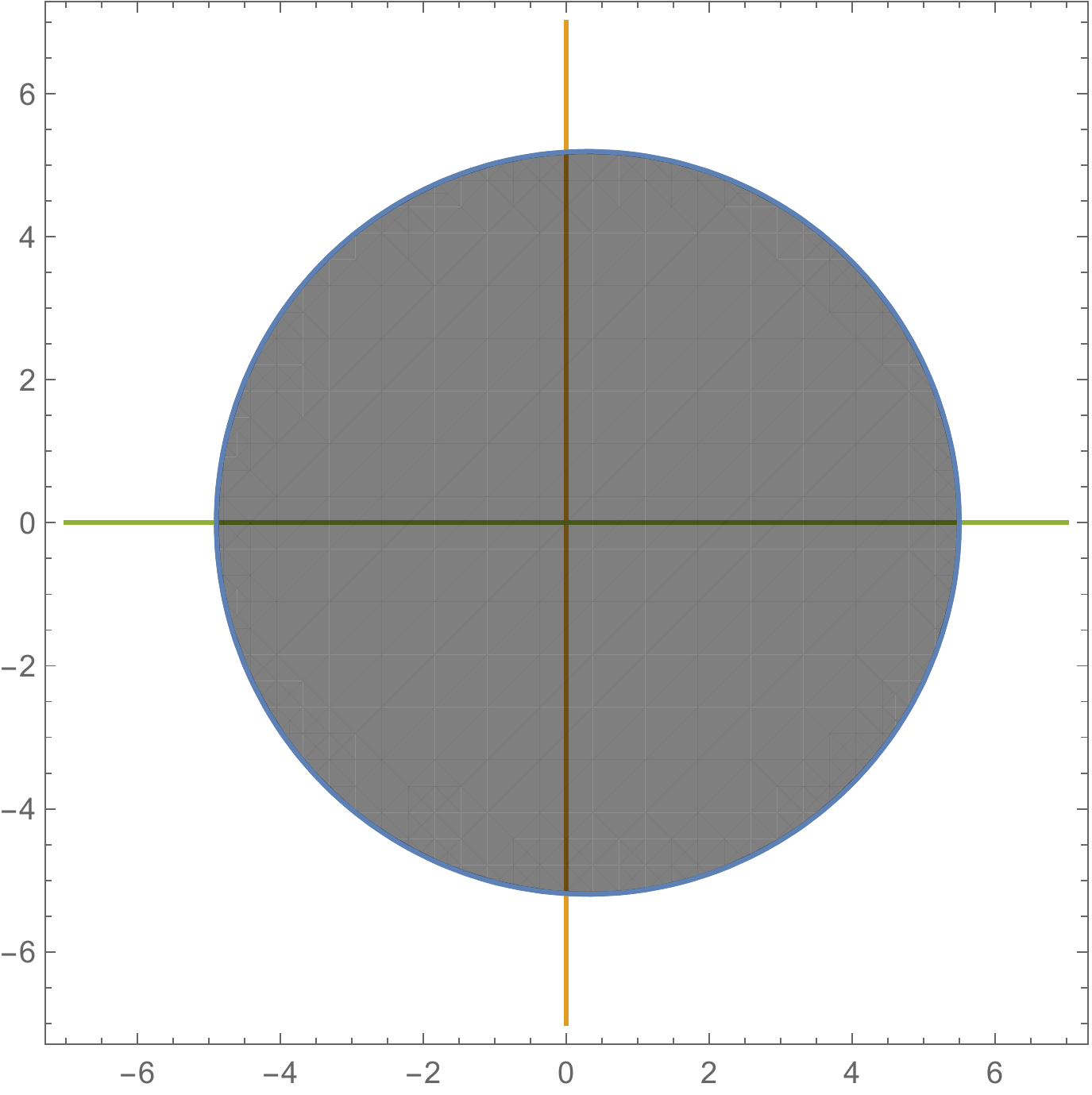}
}
\caption{Black hole shadow for small rotational rapidity parameter $\alpha = \frac{a}{M}$. For the sake of visualizing the shift of the circle's center clearly, I have taken $\alpha = 0.3$ to satisfy $\alpha \ll 1$.}
\end{figure}

\subsection{$\alpha \approx 1$ for an observer in the equatorial plane $\theta_0 = \pi/2$}
For this calculation, take the class (b) solutions for the critical impact parameters while setting $\alpha$ to unity. We get
\begin{equation}
\begin{split} 
\mathcal L_c & = - \rho_c^2 + 2 \rho_c + 1,\\
\mathcal Q_c^\circ & = \rho_c^3 (4 - \rho_c).
\end{split}
\end{equation}
Solving for $\rho_c$ inside the first equations, we get 
\begin{equation}
\rho_c = 1 + \sqrt{2 - \mathcal L_c}.
\end{equation}
Where we have safely ommited the root which is $\rho_c < 1$, since it will be inside the event horizon. \newline
Substituting this into the $\mathcal Q^\circ$ equation
\begin{equation}
\mathcal Q_c^\circ = \big( 1 + \sqrt{2 - \mathcal L_c} \big)^3 (3 - \sqrt{2 - \mathcal L_c}).
\end{equation}
Now from \eqref{coordrel}, we know $\mathcal L = - \xi^\circ$ and $\mathcal Q^\circ = \eta^{\circ \ 2}$. Finally substituting these into our last equation yields the relation between the points of the extremities of the black hole shadows
\begin{equation}
\label{firsttry}
\eta^{\circ \ 2} = \Big( 1 + \sqrt{2 + \xi^\circ} \Big)^3 \Big(3 - \sqrt{2 + \xi^\circ}\Big).
\end{equation}

\begin{figure}[H]
\centering
\subfloat[][]{
  \includegraphics[width=0.4\columnwidth]{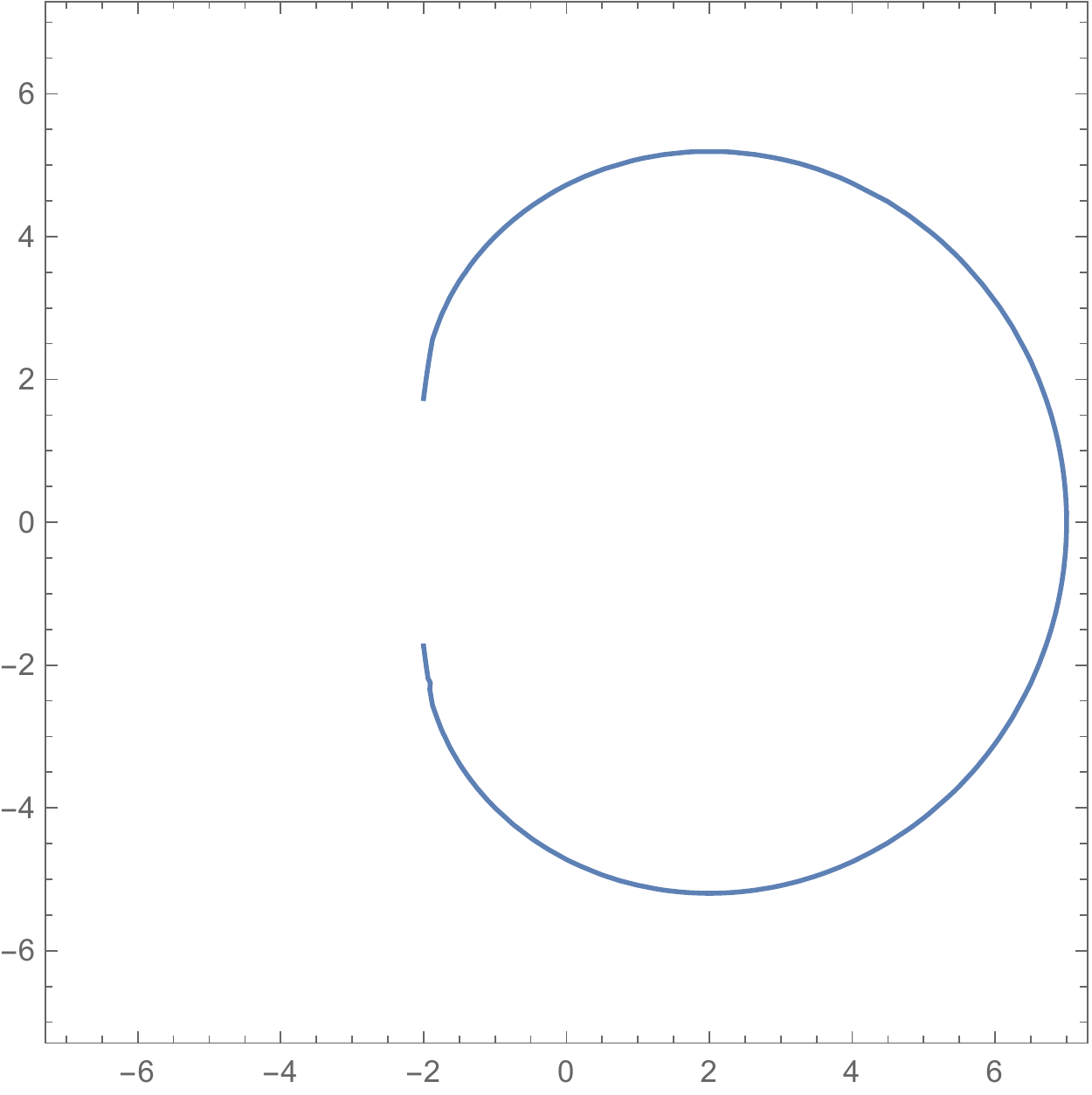}
}
\caption{The extremity of the black hole shadow as drawn from the relation \eqref{firsttry}. Note that there is a degenerate part to the left of the curve, where it is not obvious how the extermity of the shadow behaves.}
\end{figure}

So where are the outlines to the left of the figure ? Where does this shadow end ? \newline
Let us try to figure out what the curvature ($\frac{d^2 \xi^\circ}{d \eta^{\circ \ 2}}$) of the curve will be where it cuts the $\xi^\circ$ axis, to the left. As we have just shown through $\mathcal Q_c^\circ = \eta^{\circ \ 2}$, this corresponds to photons coming with parameter $\mathcal Q_c^\circ = 0$. Solving for our critical radius, knowing $\mathcal Q_c^\circ = \frac{\rho_c^3\big(4\alpha^2 - \rho_c(\rho_c-3)^2\big)}{\alpha^2(1-\rho_c)^2}$, we know that the following has to be satisfied
\begin{equation}
\begin{split} 
4\alpha^2 - \rho_c(\rho_c-3)^2 & = 0 \\
\rho_c^3  - 6 \rho_c^2 + 9 \rho_c - 4 \alpha^2 & = 0 \\
\rho_c^3  - 6 \rho_c^2 + 9 \rho_c - 4 & = 0 \qquad \qquad \text{For } \alpha \approx 1 .
\end{split}
\end{equation}
Which corresponds to a double root at $\rho_c = 1$ and a root at $\rho_c = 4$. Photons coming from the left can pass closer to the black hole, and thus will have a lower critical radius $\rho^l_c = 1$ than the ones coming from the right, which have a critical radius $\rho^r_c = 4$. Every other incoming photons coming in from different polar angles $\psi$ have a corresponding critical radius $\rho_c (\psi)$ which takes on a value strictly between $\rho^l_c < \rho_c (\psi) < \rho^r_c$.  \newline
Using the impact parameters found in terms of $\rho_c$ in section 6, it can then be checked with the help of some mathematical tool like Matlab or Mathematica that, $\frac{d^2 \xi}{d \eta^2}|_{\rho^l_c} = 0$ \cite{frolov_novikov_1998}, implying the curvature of the shadow at the left of the graph is vanishing for an extremal black hole. We can "tie things off" by drawing a straight line tying together the extermities of our graph to give us the final shadow for an extremal black hole.

\begin{figure}[H]
\centering
\subfloat[][]{
  \includegraphics[width=0.4\columnwidth]{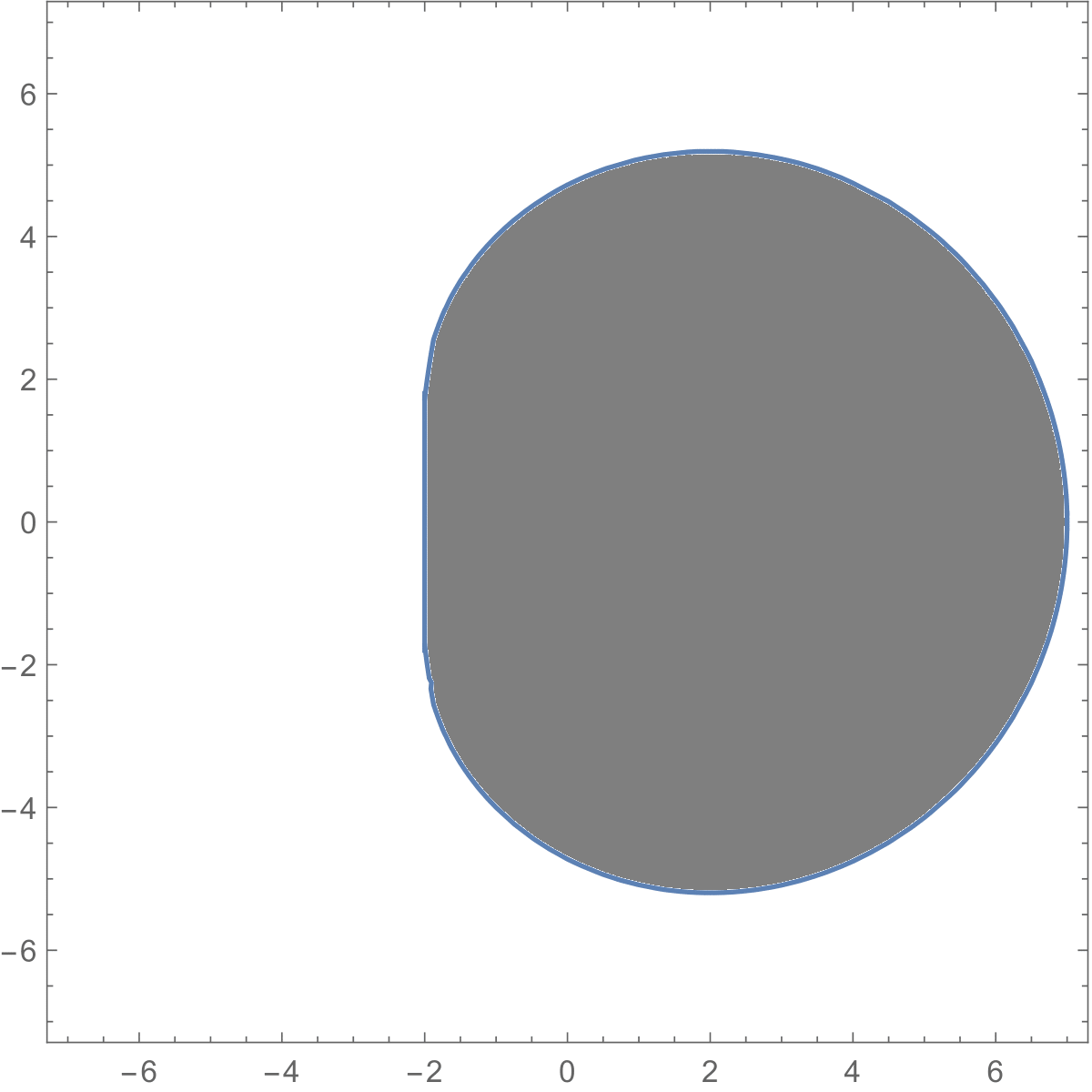}
}
\caption{The full shadow of an extremal Kerr black hole as seen by an observer at the equatorial plane.}
\end{figure}

\section{Conclusion}
We have used the properties of the Kerr spacetime to derive the conserved quantity $\mathcal Q^\circ$ throughout the motion of a particle -in this case a photon-, which then allowed us to express the equations of motion in concise, first degree equations for that particle. We have seen that the motion of such a photon is charaterized by this constant $\mathcal Q^\circ$, along with the azimuthal specific angular momentum $\mathcal L$. Through our analysis, we have uncovered that in the parameter space, there is a captured photon region of parameters for which photons having said parameters cannot escape to infinity. These allowed us to distinguish photons that can make it to an observer far away and those that cannot. We then introduced the $\textit{celestial coordinates}$ of the observer for the patch of the sky around the black hole, with the use of which we have translated the captured photon region of the parameter space into the observer's sky coordinates. Finally, we have illustrated this patch of the sky with the help of computer generated graphics to get a visualization of what we might expect to see when looking at a spinning black hole from its equatoral plane. 

\section*{Appendix A : Lagrangian formulation, Hamiltonian formulation and the Hamilton Jacobi Equation}
In classical mechanics, we define the action $S$ of a system as follows
\begin{equation}
S = \int d \tau \mathcal{L} (x,\dot{x},\tau).
\end{equation}
Where the Lagrangian $\mathcal{L}$ contains information about the dynamics of the system. We recover the equations of motion by varying this action and applying the hamilton's equations. We also have the conjugate momentum of $x^\mu$ defined as $p_\mu = \frac{\mathcal L}{\partial \dot x^\mu}$. \newline
But our lagrangian is defined it terms of $\dot x^\mu$, and not the conjugate momenta $p_\mu$ of the coordinates used. We would like to optain a quantity which encodes similar dynamics about the system, but which is given in terms of a variable which is more natural to work with. To achieve this, notice the differential form $d\mathcal{L}$ to be
\begin{equation}
\begin{split}
d\mathcal{L} & = \frac{\partial \mathcal{L}}{\partial x^\mu} d x^\mu + \frac{\partial \mathcal{L}}{\partial \dot{x}^\mu} d \dot{x^\mu} + \frac{\partial \mathcal{L}}{\partial \tau} d\tau = \frac{\partial \mathcal{L}}{\partial x^\mu} d x^\mu + p_\mu d \dot{x^\mu} + \frac{\partial \mathcal{L}}{\partial \tau} d\tau \\
& = \frac{\partial \mathcal{L}}{\partial x^\mu} d x^\mu + d (p_\mu \dot{x^\mu}) -  \dot{x^\mu} d p_\mu + \frac{\partial \mathcal{L}}{\partial \tau} d\tau, \\
d (\mathcal L - p_\mu \dot x ^\mu) & = \frac{\partial \mathcal{L}}{\partial x^\mu} d x^\mu -  \dot{x^\mu} d p_\mu + \frac{\partial \mathcal{L}}{\partial \tau} d\tau.
\end{split}
\end{equation}
Rebranding the left hand side of the above equation into a new quantity $\mathcal{H}$ defined as $\mathcal{H} \equiv p_\mu \dot x^\mu - \mathcal L$, we can note the following
\begin{equation}
d\mathcal{H} = - \frac{\partial \mathcal{L}}{\partial x^\mu} d x^\mu + \dot{x^\mu} d p_\mu - \frac{\partial \mathcal{L}}{\partial \tau} d\tau.
\end{equation}
While the following allows us to reword our equations of motion into this new language
\begin{equation}
\begin{split}
d\mathcal{H} & =  \frac{\partial \mathcal{H}}{\partial x^\mu} d x^\mu + \frac{\partial \mathcal{H}}{\partial p_\mu} d p_\mu + \frac{\partial \mathcal{H}}{\partial \tau} d\tau, \\
\frac{\partial \mathcal{H}}{\partial x^\mu} & = - \frac{\partial \mathcal{L}}{\partial x^\mu} = - \frac{d}{d \tau} \big( \frac{\partial \mathcal{L}}{\partial \dot{x}^\mu} \big) = - \dot p_\mu, \\
\frac{\partial \mathcal{H}}{\partial p_\mu} & = \dot{x^\mu}, \\
\frac{\partial \mathcal{H}}{\partial \tau} & = - \frac{\partial \mathcal{L}}{\partial \tau}.
\end{split}
\end{equation}
Where in the second line, we have made use of the Euler-Lagrange equations of motion. \newline
The lagrangian $\mathcal{L}$ and the hamiltonian $\mathcal{H}$ are said to be the Legendre transform of one another. The usefulness of hamiltonian mechanics comes from the fact that the equations obtained are of first order, and have their dynamical equations in a single, simple form. \newline
Now consider a transformation with generating function $F$ which transforms the phase space ($x^\mu$,$p_\mu$) into ($X^\mu$,$P_\mu$). This will relate the old and the new hamiltonian with
\begin{equation}
\dot x^\mu p_\mu - H = \dot X^\mu P_\mu -  K + \frac{d F}{d \tau}.
\end{equation}
Picking $F = U(x^\mu,P_\mu,\tau) - X^\mu P_\mu$
\begin{equation}
\begin{split}
\dot x^\mu p_\mu - H & = \cancelto{}{\dot X^\mu P_\mu} -  K + \frac{\partial U}{\partial x^\mu} \dot x^\mu + \frac{\partial U}{\partial P_\mu} \dot P_\mu + \frac{\partial U}{\partial \tau} \cancelto{}{- \dot X^\mu P_\mu} - X^\mu \dot P_\mu, \\
\dot x^\mu (p_\mu - \frac{\partial U}{\partial x^\mu}) - H & = \dot P_\mu (\frac{\partial U}{\partial P_\mu} - X^\mu) - K  + \frac{\partial U}{\partial \tau}.
\end{split}
\end{equation}
Since ($x^\mu$,$p_\mu$) and ($X^\mu$,$P_\mu$) are taken to be independent, we know
\begin{equation}
\begin{split}
\label{GenFunc}
p_\mu & = \frac{\partial U}{\partial x^\mu}, \\
X^\mu & = \frac{\partial U}{\partial P_\mu}, \\
K & = H + \frac{\partial U}{\partial \tau}.
\end{split}
\end{equation}
In addition, we would like our new coordinates to be cyclic. Meaning that in our new formulation,
\begin{equation}
\begin{split}
\frac{\partial K}{\partial X^\mu} & = - \dot P_\mu = 0, \\
\frac{\partial K}{\partial P_\mu} & = \dot X^\mu = 0.
\end{split}
\end{equation}
Hinting at the constancy of $K$ with respect to these new coordinates. We can pick $K=0$ without loss of generality. \newline
From \eqref{GenFunc}, we then get
\begin{equation}
\label{HJEqu}
K = H + \frac{\partial U}{\partial \tau} = 0 \rightarrow \frac{\partial U}{\partial \tau} = \frac{m}{2}.
\end{equation}
Along with an anzats to the function $U$
\begin{equation}
U = \frac{m}{2} \tau - E t + L \phi + U_r (r) + U_\theta (\theta).
\end{equation}
Where, as seen from \eqref{GenFunc}
\begin{equation}
\begin{split}
p_r & = \frac{\partial U_r (r)}{\partial r}, \\
p_\theta & = \frac{\partial U_\theta (\theta)}{\partial \theta}.
\end{split}
\end{equation}

\section*{Appendix B : Free test particle lagrangian and equations of motion}
Consider a test particle of mass m in an arbitrary metric $g_{\mu \nu}$ - we specified \textit{test particle} to infer the particle's contribution to the curving of spacetime is negligible -. The lagrangian for such a system can be written in the form
\begin{equation}
\mathcal{L}_m = \frac{m}{2}g_{\mu \nu} (x) \frac{d x^\mu}{d \tau} \frac{d x^\nu}{d \tau}.
\end{equation}
\begin{equation}
\begin{split}
S & = \frac{m}{2} \int d \tau \ g_{\mu \nu} (x) \frac{d x^\mu}{d \tau} \frac{d x^\nu}{d \tau}, \\
\delta S & =  \frac{m}{2} \int d\lambda \ \big( \frac{d x^\mu}{d \tau} \frac{d x^\nu}{d \tau} \delta g_{\mu \nu} (x) + 2 g_{\mu \nu} (x) \frac{d x^\mu}{d \tau} \delta \frac{d x^\nu}{d \tau} \big) \\
& = m \int d\lambda \ \big( \frac{1}{2} \frac{d x^\mu}{d \tau} \frac{d x^\nu}{d \tau} \partial_\rho g_{\mu \nu} \delta x^\rho + g_{\mu \nu} (x) \frac{d x^\mu}{d \tau} \frac{d }{d \tau} \delta x^\nu \big) \\
& = m \int d\lambda \Big\{ \frac{d}{d \tau} \big( g_{\mu \nu} \frac{dx^\mu}{d \tau} \delta x^\nu \big) - \big( g_{\mu \nu} \frac{d^2 x^\mu}{d \tau ^2} + \frac{dx^\mu}{d \tau}\frac{dx^\rho}{d \tau} \partial_\rho g_{\mu \nu} - \frac{1}{2} \frac{d x^\mu}{d \tau} \frac{d x^\rho}{d \tau} \partial_\nu g_{\mu \rho} \big) \delta x^\nu \Big\} \\
& = m \int d\lambda \Big\{ \frac{d}{d \tau} \big( g_{\mu \nu} \frac{dx^\mu}{d \tau} \delta x^\nu \big) - \big(\frac{d^2 x^\sigma}{d \tau ^2} + \underbrace{\frac{1}{2} g^{\nu \sigma} (\partial_\rho g_{\mu \nu} + \partial_\mu g_{\rho \nu}-  \partial_\nu g_{\mu \rho})}_{\Gamma^\sigma_{\ \mu \rho}} \frac{dx^\mu}{d \tau}\frac{dx^\rho}{d \tau} \big)  \delta x_\sigma \Big\} \\
& = m \int d\lambda \Big\{ \frac{d}{d \tau} \big( g_{\mu \nu} \frac{dx^\mu}{d \tau} \delta x^\nu \big) - \big(\frac{d^2 x^\sigma}{d \tau ^2} + \Gamma^\sigma_{\ \mu \rho} \frac{dx^\mu}{d \tau}\frac{dx^\rho}{d \tau} \big)  \delta x_\sigma \Big\}. \\
\end{split}
\end{equation}
The boundary term vanishes since the variation $\delta x^\nu$ is taken to be vanishing at the boundaries, whilst the second integrand should satisfy Hamilton's principle for arbitrary values of $\delta x_\sigma$ throughout the motion. We recover the geodesic equation :
\begin{equation}
\frac{D^2 x^\sigma }{d \tau ^2}= \frac{d^2 x^\sigma}{d \tau ^2} + \Gamma^\sigma_{\ \mu \rho} \frac{dx^\mu}{d \tau}\frac{dx^\rho}{d \tau} = 0.
\end{equation}

\section*{Appendix C : Lie derivative and Killing vector fields}
A vector field $v^\mu$ defined everywhere on the manifold can help us quantify some infinitesimal transformation along that vector field of each point $x^\mu$ to a new nearby point $x'^\mu$
\begin{equation}
x^\mu \rightarrow x'^\mu = x^\mu + \epsilon v^\mu(x) \qquad \qquad \text{Where } \epsilon \text{ is some arbitrarily small parameter.}
\end{equation}
Under such a coordinate transformation, it is easy to check that a scalar field $\phi(x)$ defined on the manifold transforms as
\begin{equation}
\begin{split}
\phi(x) \rightarrow \phi(x') & = \phi(x+\epsilon v) = \phi(x) + \epsilon v^\mu \partial_\mu \phi.
\end{split}
\end{equation}
The quantity of change of this scalar field under such a transformation along $v^\mu$ is said to be the \textit{Lie Derivative} of $\phi$ by $v$, and is expressed as
\begin{equation}
\mathcal L_v \phi = v^\mu \partial_\mu \phi.
\end{equation}
We can extend the definition of Lie Derivatives to covariant vector fields, say, $\omega_\mu$. Using the invariance of the combination $\omega_\mu dx^\mu$ (This implies $\omega_\mu (x) dx^\mu$ will strictly go to $\omega_\mu (x') dx'^\mu$ after the transformation)
\begin{equation}
\begin{split}
\omega_\mu (x) dx^\mu \rightarrow \omega_\mu (x') dx'^\mu & = \omega_\mu (x+\epsilon v ) \big( dx^\mu + \epsilon (\partial_\alpha v^\mu) dx^\alpha \big) \\
& = \big( \omega_\mu (x) + \epsilon \partial_\alpha \omega_\mu v^\alpha \big) \big( dx^\mu + \epsilon (\partial_\alpha v^\mu) dx^\alpha \big) \\
& = \omega_\mu (x) dx^\mu + \epsilon \big( v^\alpha \partial_\alpha \omega_\mu dx^\mu + w_\mu \partial_\alpha v^\mu dx^\alpha \big) + \mathcal O (\epsilon ^2).
\end{split}
\end{equation}
From which we obtain
\begin{equation}
\mathcal L_v \omega_\mu = v^\alpha \partial_\alpha \omega_\mu + \omega_\alpha \partial_\mu v^\alpha.
\end{equation}
One can similarly calculate the Lie Derivative of a second rank covariant tensor (say, the metric tensor two form components $g_{\mu \nu}$).
\begin{equation}
\mathcal L_v g_{\mu \nu} = v^\alpha \partial_\alpha g_{\mu \nu} + g_{\alpha \nu} \partial_\mu v^\alpha + g_{\mu \alpha} \partial_\nu v^\alpha.
\end{equation}
Which can be reexpressed as follows
\begin{equation}
\begin{split}
\mathcal L_v g_{\mu \nu} & = v^\alpha \partial_\alpha g_{\mu \nu} + g_{\alpha \nu} \partial_\mu v^\alpha + g_{\mu \alpha} \partial_\nu v^\alpha = \underbrace{v^\alpha \partial_\alpha g_{\mu \nu} - v^\alpha \partial_\mu g_{\alpha \nu} - v^\alpha \partial_\nu g_{\mu \alpha}}_{2 v_\alpha \Gamma^\alpha_{\ \mu \nu}} + \partial_\mu v_\nu + \partial_\nu v_\mu \\
& = \nabla_\mu v_\nu + \nabla_\nu v_\mu.
\end{split}
\end{equation}
Any vector field $v^\mu$ which leaves the metric components unchanged along itself is called \textit{Killing vector field}. We will denote them as $K^\mu$. It is straightforwards from the result we have just found above that a killing vector field $K^\mu$ satisfies
\begin{equation}
\nabla_\mu K_\nu + \nabla_\nu K_\mu = 0.
\end{equation}

\section*{Appendix D : Conserved charge from Killing vector fields}
Let us now inquire the result of the following total proper time derivative, where $K^\mu$ is a Killing vector field and $x^\mu$ are geodesics.
\begin{equation}
\begin{split}
\frac{d}{d \tau} \big( g_{\mu \nu} K^\mu \frac{d x^\nu}{d \tau}\big) & = K^\mu \frac{d}{d \tau} \big( g_{\mu \nu} \frac{d x^\nu}{d \tau}\big) + g_{\mu \nu} \frac{d K^\mu}{d \tau} \frac{d x^\nu}{d \tau} \\
& =  K^\mu \frac{d}{d \tau} \big( g_{\mu \nu} \frac{d x^\nu}{d \tau}\big) + g_{\mu \nu} \partial_\alpha K^\mu \frac{d x^\alpha}{d \tau} \frac{d x^\nu}{d \tau}.
\end{split}
\end{equation}
Where we can work on the $\frac{d}{d \tau} \big( g_{\mu \nu} \frac{d x^\nu}{d \tau}\big)$ term using the geodesic equation (Check Appendix B)
\begin{equation}
\begin{split}
\frac{d}{d \tau} \big( g_{\mu \nu} \frac{d x^\nu}{d \tau}\big) & = \frac{d g_{\mu \nu}}{d \tau} \frac{d x^\nu}{d \tau} + g_{\mu \nu} \frac{d^2 x^\nu}{d \tau^2} = \partial_\alpha g_{\mu \nu} \frac{d x^\alpha}{d \tau} \frac{d x^\nu}{d \tau} - g_{\mu \nu} \Gamma^\nu_{\ \alpha \beta} \frac{d x^\alpha}{d \tau} \frac{d x^\beta}{d \tau} \\
& = \big( \partial_\alpha g_{\mu \beta} - \frac{1}{2} (\partial_\alpha g_{\mu \beta} + \partial_\beta g_{\mu \alpha} - \partial_\mu g_{\alpha \beta}) \big) \frac{d x^\alpha}{d \tau} \frac{d x^\beta}{d \tau} \\
& = \big( \partial_\alpha g_{\mu \beta} - \partial_\alpha g_{\mu \beta} + \frac{1}{2} \partial_\mu g_{\alpha \beta}) \big) \frac{d x^\alpha}{d \tau} \frac{d x^\beta}{d \tau} \\
& = \frac{1}{2} \partial_\mu g_{\alpha \beta} \frac{d x^\alpha}{d \tau} \frac{d x^\beta}{d \tau}.
\end{split}
\end{equation}
Where on the fourth line, we have made use of the symmetric nature of the $\alpha \leftrightarrow \beta$ indices. \newline
Returning to the quantity we were evaluating
\begin{equation}
\begin{split}
\frac{d}{d \tau} \big( g_{\mu \nu} K^\mu \frac{d x^\nu}{d \tau}\big) & = \frac{1}{2} K^\mu \partial_\mu g_{\alpha \beta} \frac{d x^\alpha}{d \tau} \frac{d x^\beta}{d \tau} + g_{\mu \nu} \partial_\alpha K^\mu \frac{d x^\alpha}{d \tau} \frac{d x^\nu}{d \tau} \\
& = \frac{1}{2} \big( \underbrace{ K^\mu \partial_\mu g_{\alpha \beta} + g_{\mu \beta} \partial_\alpha K^\mu + g_{\alpha \mu} \partial_\beta K^\mu}_{\mathcal L_K g_{\alpha \beta}} \big) \frac{d x^\alpha}{d \tau} \frac{d x^\beta}{d \tau} \\
& = \frac{1}{2} \mathcal L_K g_{\alpha \beta} \frac{d x^\alpha}{d \tau} \frac{d x^\beta}{d \tau} = 0.
\end{split}
\end{equation}
Where on the second line, we have yet again made use of the symmetry in the $\alpha \leftrightarrow \beta$ indices. \newline
Along a Killing vector field, the metric's Lie derivative is identically zero. Thus, the quantity $g_{\mu \nu} K^\mu \frac{dx^\nu}{d \tau}$ is conserved.

\section*{Acknowledgments}
I would like to thank Prof. Dr. Bayram Tekin for his answers to my numerous questions and comments on various parts of this paper. \newline
Special thanks to my family for always supporting me through any of my endeavors.

\end{document}